\newif\ifLINUXBUILD
\newif\ifAPPENDIX
\IfFileExists{/dev/null}{\LINUXBUILDtrue}{}
\documentclass[sigconf,nonacm=true]{acmart}

\author{Sohaib ul Hassan}
\affiliation{
\institution{Tampere University}
\city{Tampere}
\country{Finland}
}
\email{sohaibulhassan@tuni.fi}

\author{Iaroslav Gridin}
\orcid{0000-0002-1239-1841}
\affiliation{
\institution{Tampere University}
\city{Tampere}
\country{Finland}
}
\email{iaroslav.gridin@tuni.fi}

\author{Ignacio M. Delgado-Lozano}
\orcid{0000-0002-0003-3318}
\affiliation{
\institution{Tampere University}
\city{Tampere}
\country{Finland}
}
\email{ignacio.delgadolozano@tuni.fi}

\author{Cesar Pereida Garc\'ia}
\orcid{0000-0001-6812-8498}
\affiliation{
\institution{Tampere University}
\city{Tampere}
\country{Finland}
}
\email{cesar.pereidagarcia@tuni.fi}

\author{Jes\'us-Javier Chi-Dom\'inguez}
\orcid{0000-0002-9753-7263}
\affiliation{
\institution{Tampere University}
\city{Tampere}
\country{Finland}
}
\email{jesus.chidominguez@tuni.fi}

\author{Alejandro Cabrera Aldaya}
\orcid{orcid.org/0000-0002-1544-6772}
\affiliation{
\institution{Tampere University}
\city{Tampere}
\country{Finland}
}
\email{alejandro.cabreraaldaya@tuni.fi}

\author{Billy Bob Brumley}
\orcid{0000-0001-9160-0463}
\affiliation{
\institution{Tampere University}
\city{Tampere}
\country{Finland}
}
\email{billy.brumley@tuni.fi}

\usepackage[T1]{fontenc}
\usepackage{graphicx}
\usepackage{xspace}
\usepackage{lipsum}
\usepackage[linesnumbered, ruled, vlined]{algorithm2e}
\usepackage{mathtools}
\usepackage{multirow}
\DeclarePairedDelimiter{\ceil}{\lceil}{\rceil}

\newcommand{\Paragraph}[1]{\subsubsection*{#1}}

\DeclareMathOperator{\lcm}{lcm}

\newcommand{\CVE}[1]{\href{https://cve.mitre.org/cgi-bin/cvename.cgi?name=CVE-#1}{\mbox{CVE-#1}}}

\newcommand{\footurl}[1]{\footnote{\url{#1}}}
\newcommand{\rfc}[1]{\href{https://tools.ietf.org/html/rfc#1}{RFC #1}~\citep{rfc:#1}}

\newcommand{\add}{\emph{add}\xspace}
\newcommand{\dbl}{\emph{double}\xspace}

\newcommand{\Oinf}{\ensuremath{\mathcal{O}}\xspace}

\newcommand{\ie}{i.e.,\xspace}
\newcommand{\eg}{e.g.,\xspace}
\newcommand{\cf}{cf.\,}

\newcommand{\wnaf}{wNAF\xspace}
\newcommand{\code}[1]{\texttt{#1}\xspace}
\newcommand{\ecc}[1]{\texttt{#1}\xspace}
\newcommand{\NSS}{NSS\xspace}

\newcommand{\Libwnaf}{\code{ec\_compute\_wNAF}}

\newcommand{\LibIsOdd}{\code{mp\_isodd}}
\newcommand{\LibSub}{\code{mp\_sub\_d}}
\newcommand{\LibAdd}{\code{mp\_add\_d}}
\newcommand{\LibCmpDig }{\code{s\_mp\_cmp\_d}}
\newcommand{\graphene}{\code{Graphene-SGX}}

\newcommand{\nssVer}{v3.51\xspace}
\newcommand{\PKCSONCE}{PKCS \#11\xspace}
\newcommand{\cveRemoteStill}{\CVE{2011-1945}\xspace}

\newcommand{\twodots}{\mathinner {\ldotp \ldotp}}

\newcommand{\sgxstep}{\code{SGX-Step}}
 
\newcommand{\KEYWORDS}{%
applied cryptography;
public key cryptography;
DSA;
ECDSA;
RSA;
side-channel analysis;
lattice-based cryptanalysis;
software security;
NSS;
CVE-2020-12399;
CVE-2020-12402;
CVE-2020-6829;
CVE-2020-12401}

\title{D\'{e}j\`{a} Vu: Side-Channel Analysis of Mozilla's NSS}

\begin{abstract}
Recent work on Side Channel Analysis (SCA) targets old, well-known
vulnerabilities, even previously exploited, reported, and patched in
high-profile cryptography libraries. Nevertheless, researchers continue to find and exploit
the same vulnerabilities in old and new products, highlighting a big issue among
vendors: effectively tracking and fixing security vulnerabilities when
disclosure is not done directly to them.
In this work, we present another instance of this issue by performing the first
library-wide SCA security evaluation of Mozilla's NSS security library.
We use a combination of two independently-developed SCA security
frameworks to identify and test security vulnerabilities.
Our evaluation uncovers several new vulnerabilities in NSS
affecting DSA, ECDSA, and RSA cryptosystems. We exploit said vulnerabilities and
implement key recovery attacks using signals---extracted through different techniques
such as timing, microarchitecture, and EM---and improved lattice methods.
 \end{abstract}
\keywords{\KEYWORDS{}}

\begin{document}

\maketitle

\section{Introduction} \label{sec:intro}

Traditionally, SCA security research involves manual analysis
of software libraries by actively triggering the execution paths of
cryptographic schemes in isolation, and measuring information leaks.
Constant-time software implementation is considered the most widely
adapted countermeasure against these information leaks
\cite{DBLP:conf/acsac/TuveriHGB18}. However, owing to a long list of
SCA vulnerabilities in popular software libraries such as
OpenSSL \cite{DBLP:conf/esorics/BrumleyT11,
DBLP:conf/ches/BengerPSY14, DBLP:conf/ccs/GarciaBY16,
DBLP:conf/uss/GarciaB17, DBLP:conf/sp/AldayaBHGT19}, we argue that for
cryptographic library maintainers and developers, manual verification
for constant-time behavior is a non-trivial task, and requires
extensive knowledge about SCA security. This can result in
SCA vulnerabilities being overlooked by \emph{peer vendors},
especially when the issues are not directly reported to them.

A peer vendor is one that is ``at the same horizontal level of the supply
chain; peer vendors may be independent implementers of the same technology
(\eg OpenSSL and GnuTLS)'' \cite[p.\ 4]{2017:first}.
In the context of our work, \NSS, OpenSSL, GnuTLS, BoringSSL, LibreSSL, mbedTLS,
WolfSSL, etc.\ all fit this definition of peer vendors,
sharing the same market vertical, implementing (at least) several versions of
the TLS standard and many of the implied cryptographic algorithms.

Inspired by documented multi-vendor security failures (detailed later in
\autoref{sec:peer}), in this work we analyze the SCA
security of Mozilla's \NSS. OpenSSL's rich history provides a large corpus of
security vulnerabilities, many of which are root caused to failure to use
constant-time algorithms. We leverage this corpus to explore how well \NSS has
kept up with OpenSSL's significant improvements that accelerated after
HeartBleed (\CVE{2014-0160}).

Taking NSS as our case study, we present a novel approach to SCA
security, by developing a systematic methodology for library-wide
automated identification of SCA leaks and flagging the
vulnerable execution paths. Our findings reveal serious SCA
deficiencies in NSS, which not only questions the current practice of
vulnerability disclosure among peer vendors, but more importantly provide a
general approach to automated SCA security validation for
cryptographic library developers. Furthermore, we also perform
multiple end-end attacks to highlight the severity of the discovered
vulnerabilities and responsibly assist Mozilla to mitigate them.

\Paragraph{Contributions}
Briefly, the contributions of our work include:
(i) combining the DATA \cite{DBLP:conf/uss/WeiserZSMMS18} and TriggerFlow
\cite{DBLP:conf/dimva/GridinGTB19} frameworks to close the gap between identifying
leakage and software / module / unit / regression testing, subsequently
applied to \NSS (\autoref{sec:vulns}) to discover, test, and exploit several
novel vulnerabilities;
(ii) an end-to-end network-based remote timing attack against \NSS's DSA
implementation (\autoref{sec:attack_dsa});
(iii) discovery of a traditional timing attack vulnerability in \NSS's ECDSA
implementation (\autoref{sec:attack_padding}) in the context of nonce padding;
(iv) an end-to-end ElectroMagnetic Analysis (EMA) attack on \NSS's ECDSA implementation
(\autoref{sec:attack_wnaf}) in the context of Elliptic Curve Cryptography (ECC)
scalar multiplication;
(v) an end-to-end microarchitecture attack on \NSS's ECDSA implementation
(\autoref{sec:attack_sgx}) in the context of ECC scalar recoding;
(vi) an end-to-end EMA attack on \NSS's RSA key generation implementation
(\autoref{sec:attack_rsa});
(vii) improved lattice methods and empirical data for realizing several of these
end-to-end attacks (\autoref{sec:lattice}).
\section{Background} \label{sec:background}

\subsection{Public Key Cryptography}

\Paragraph{DSA}
Denote primes $p,q$ such that $q$ divides $(p-1)$, and a generator $g \in GF(p)$ of
multiplicative order $q$.
The user's private key $\alpha$ is an integer uniformly chosen from
$\{1 \twodots q-1\}$ and the corresponding public key is
$y = g^{\alpha} \bmod p$.
With approved hash function $\textrm{Hash}()$, the DSA digital
signature $(r,s)$ on message $m$ (denoting with $h < q$ the
representation of $ \textrm{Hash}(m)$ as an integer) is
\begin{equation} \label{eq:dsa}
    r = (g^{k} \bmod p) \bmod q, \quad s = k^{-1} (h + \alpha r) \bmod q
\end{equation}
where $k$ is a nonce chosen uniformly from $\{1 \twodots q-1\}$.

\Paragraph{ECDSA}
Denote an order-$q$ generator $G \in E$ of an elliptic curve group $E$ with
cardinality $fq$ and $q$ a large prime and $f$ the small cofactor.
The user's private key $\alpha$ is an integer uniformly chosen from
$\{1 \twodots q-1\}$ and the corresponding public key is $D = [\alpha]G$.
With approved hash function $\textrm{Hash}()$, the ECDSA digital
signature $(r,s)$ on message $m$ (denoting with $h < q$ the
representation of $ \textrm{Hash}(m)$ as an integer) is
\begin{equation} \label{eq:ecdsa}
r = ([k]G)_x \bmod q, \quad s = k^{-1} (h + \alpha r) \bmod q
\end{equation}
where $k$ is a nonce chosen uniformly from $\{1 \twodots q-1\}$.

\Paragraph{Point multiplication}
During ECDSA signing, point multiplication $[k]G$
is the most computationally intensive part. For curves over
prime fields, windowed non-adjacent form (\wnaf) is
a textbook method for performing point multiplication. Given a
window size $w$, and a set of pre-computed points
$\pm G,\pm[3]G, \cdots ,\pm[2^{w-1}-1]G$, the $\ell$-bit scalar can be
recoded as
\[
k = \sum_{i=0}^{\ell} k_i 2^i \qquad \mbox{where } k_i \in \{0, \pm 1, \pm 3, \cdots, \pm (2^{w-1}-1) \}
\]
The \wnaf point multiplication method (\autoref{alg:wnaf_scalar_mult})
scans signed $k_i$ digits performing a point \dbl at each step,
whereas the position of non-zero $k_i$ decides on point \add.
The \wnaf representation (\autoref{alg:compute_wanf}) reduces the number of non-zero scalar digits
to about $\ell / (w+1)$, resulting in less point additions since it
guarantees at most one out of $w$ consecutive digits are non-zero.

\ifAPPENDIX\else
\begin{algorithm}[h]
	\caption{Compute \wnaf representation of $k$}\label{alg:compute_wanf}
	\DontPrintSemicolon
	\KwIn{Integer $k$ and width $w$}
	\KwOut{$wNAF(k,w)$}
	\SetKw{KwDownTo}{downto}
	\SetKwFunction{odd}{odd}
	$i=0$\\
	\While{$k \ne 0$}
	{
		\If{$\odd(k)$}{
			$d = k \bmod 2^w$\\
			\lIf{$d > 2^{w-1}$}
			{
				$d = d - 2^w$
			}
			$k = k - d$
		}
		\lElse
		{
			$d = 0$
		}
		$wNAF[i] = d$, $k = k / 2$, $i = i + 1$
	}
	\Return{$wNAF$}
\end{algorithm}
\fi

\ifAPPENDIX\else
\begin{algorithm}[h]
	\caption{\wnaf-based scalar multiplication}\label{alg:wnaf_scalar_mult}
	\DontPrintSemicolon
	\KwIn{Integer $k$, width $w$ and elliptic curve point $G$}
	\KwOut{$R=kG$}
	\SetKw{KwDownTo}{downto}
	Compute $wNAF(k,w)$ using \autoref{alg:compute_wanf}\\
	$P[i] = iG$ ; $i \in \{-2^{w-1} + 1, \twodots, -3, -1, 1,  \twodots, 2^{w-1} - 1\}$\\
	$R = \Oinf$\\
	\For{$i = \lfloor \lg(k) \rfloor + 1$ \KwDownTo $0$}{
		$R = 2R$ \\
		\lIf{$wNAF[i] \ne 0$}{%
			$R = R + P[wNAF[i]]$}%
	}
	\Return{$R$}
\end{algorithm}
\fi

\Paragraph{RSA}
According to PKCS~\#1\xspace v2.2 (\rfc{8017}),
an RSA private key consists of the eight parameters
$\{N, e, p, q, d, d_p, d_q, i_q \}$
where all but the first two are secret, and $N=pq$ for primes $p$, $q$.
Public exponent $e$ is usually small and the following holds:
\begin{equation} \label{eq:d}
d = e^{-1} \bmod \lcm(p-1, q-1)
\end{equation}
In addition, Chinese Remainder Theorem (CRT) parameters are stored for speeding
up RSA computations:
\begin{equation} \label{eq:crt_parameters}
d_p = d \bmod p, \quad d_q = d \bmod q, \quad i_q = q^{-1} \bmod p
\end{equation}
Currently, the minimum recommended length for $N$ is 2048 bits (\ie $p$ and
$q$ are 1024-bit primes) and $e$ is fixed to a small value, typically 65537.
Regarding RSA security, beyond traditional factoring \citet{DBLP:conf/eurocrypt/Coppersmith96}
showed if we know half of the bits of either $p$ or $q$ it is possible to factor $N$
in polynomial time, a fact we will utilize later in \autoref{sec:attack_rsa}.

During key generation, a typical check is coprimality of $e$ with both $p-1$ and
$q-1$ often implemented with the binary extended Euclidean algorithm (BEEA)
\cite{1967:Stein}.
\autoref{alg:BEEA_alg} (and variants) computes the GCD of two integers $a$ and
$b$ employing solely right-shift operations (SHIFTS) and subtractions (SUBS)
instead of divisions.
The BEEA control flow strongly depends on its inputs, and an SCA capable
attacker differentiating between SUBS and SHIFT operations can recover
information on
$a$ and $b$ \cite{DBLP:journals/tches/AldayaGTB19,DBLP:conf/ccs/WeiserSB18,DBLP:conf/uss/GarciaB17,DBLP:journals/jce/AldayaSS17,DBLP:conf/ima/AciicmezGS07}.

In the context of RSA, the integer arguments will be $e$ (public) and $p-1$ or
$q-1$, putting keys at risk. It is important to note that step 4 will never
execute since $e$ is always an odd number, and that during the first iterations
$v > u$ as $p,q \gg e$.

\ifAPPENDIX\else
\begin{algorithm}[h]
\caption{Binary extended Euclidean algorithm (BEEA)}
\label{alg:BEEA_alg}
\DontPrintSemicolon
\KwIn{Integers $a$ and $b$ such that $0<a<b$}
\KwOut{Greatest common divisor of $a$ and $b$}
\SetKwFunction{even}{even}
\Begin{
    $u \gets a$, $v \gets b$, $i \gets 0$\;
    \While{\even{$u$} {\bf and} \even{$v$}}{
        $u \gets u / 2$,
        $v \gets v / 2$,
        $i \gets i + 1$
    }
    \While{$u \ne 0$}{
        \lWhile{\even{$u$}}{
            $u \gets u / 2$
        }
        \lWhile{\even{$v$}}{
            $v \gets v / 2$\
        }
        \lIf { $u \ge v$}{
            $u \gets u - v$
        }\lElse{
            $v \gets v - u$
        }
    }
    \Return{$v \cdot 2^i$} \label{algstep:gcd_return}
}
\end{algorithm}
\fi

\subsection{Peer Vendors and Security Disclosure} \label{sec:peer}

Multi-vendor security
disclosure is a practical challenge with peer vendor competing products. Quoting
\cite[p.\ 20]{2017:first}: ``Missing or poor communication between peer vendors
can negatively impact coordination efforts.'' The document goes on to give two
illustrative examples of security failures due to lack of peer vendor
communication, which we extend with a third.

\Paragraph{HTTP proxy poisoning}
First publicly disclosed by R.\,L.\ Schwartz in 2001, by setting the (undefined)
\texttt{Proxy} header in an HTTP request to a malicious URL, per \rfc{3875}
server-side CGI scripts will translate this to the \texttt{HTTP\_PROXY}
environmental variable. This is unfortunately a common environmental variable
used by subsequent server-side scripts to configure outgoing proxy connections,
redirecting traffic to the malicious URL. Rediscovered several times since, in
2016 ``httpoxy'' by D.\ Scheirlinck et al.\ led to over 14 CVE assignments
across different vendors\footurl{https://httpoxy.org/}.

\Paragraph{DNS cache poisoning}
In 1999, D.\,J.\ Bernstein implemented UDP source port randomization to harden
dnscache (part of djbdns) against DNS spoofing attacks\footurl{https://cr.yp.to/djbdns/forgery.html}.
Around that time, Bernstein made a very clear and public argument that
transaction ID randomization was insufficient. In 2008, D.\ Kaminsky developed
an exploit around the concept\footurl{https://dankaminsky.com/2008/07/09/an-astonishing-collaboration/},
resulting in \CVE{2008-1447} that affected several widely-deployed DNS solutions
at the time, such as BIND and Windows DNS.

\Paragraph{Bit lengths can be secrets, too}
In 2011, \citet{DBLP:conf/esorics/BrumleyT11} described a timing vulnerability
present in OpenSSL's implementation of ECDSA that used Montgomery's ladder as a
scalar multiplication algorithm for binary curves, exploited to steal the secret
key of a remote TLS server. The general message from the work was that in
nonce-based digital signature schemes, the effective bit length of the nonce
must also be kept secret. OpenSSL responded by issuing \CVE{2011-1945} and peer
vendor Mozilla ported the patch that landed in OpenSSL to
\NSS\footurl{https://hg.mozilla.org/projects/nss/rev/079cfc4710c7193ef73888394f4d4f935e03f241}.
In 2019, ``Minerva'' by \citet{temp:minerva} revealed the vulnerability
persists in many modern implementations leading to at least seven CVE
assignments across vendors, and similarly the recent ``TPM-FAIL'' by
\citet{temp:tmpfail} with more of a hardware focus and CVE assignments by Intel
(\CVE{2019-11090}) and STMicroelectronics (\CVE{2019-16863}).

The three examples above are an illustrative range of porting attack concepts to
related products that implement similar standards. The first being closer to the
traditional software security side, the second more of an algorithmic attack,
and the third (strongly motivating our work) a side-channel attack.
At a high level, the root cause of all
three examples is vendors failing to follow the advice of security professionals
and being blind to peer vendor activity regarding security hardening.

\subsection{Mozilla Network Security Services (NSS)}

With its roots in Netscape Navigator and SSLv2, \NSS is a Free and Open Source Software (FOSS) project tailored
for Internet security and interoperating security (\eg TLS) and crypto (\eg
\PKCSONCE) standards. The software suite consists of two libraries
(\code{libnss} along with Netscape Portable Runtime \code{libnspr}) and over 70
CLI tools linking against them---\eg certutil, pk1sign, p7sign, p7verify,
signtool, etc.

Mozilla maintains the development infrastructure for \NSS and is the main
contributor due to the history of the project, as well as Firefox's significant
browser market share. Yet currently Red Hat is also a major contributor due to
their server-side enterprise use cases, and over the years other contributors
include Sun Microsystems/Oracle Corporation, Google, and AOL.

Below the TLS layer, what differentiates \NSS from other crypto-featured security
software libraries is its abstraction of cryptographic operations. It features
native \PKCSONCE support for hardware and software security modules. In fact, at
the API level, this is the interface at which linking applications drive the
crypto---\NSS serves crypto operations backed either externally through a
\PKCSONCE hardware token, or (by default) internally through its own \PKCSONCE
software token. This is different from all other major libraries, \eg OpenSSL
which provides access to crypto functionality either through its EVP interface
(modern) or directly through low level APIs (legacy).

The comparison between \NSS and OpenSSL is important due to the shared history
and evolution of the projects. In particular, starting in 2001 Oracle (then Sun
Microsystems) made fundamental FOSS contributions by integrating their ECC
software into both projects\footurl{https://seclists.org/isn/2002/Sep/89}, a new
feature for both libraries. In that respect, the ECC parts of both libraries are
forked from the same original code \cite{DBLP:conf/www/GuptaSS04}, yet have
evolved independently over the last two decades.

\Paragraph{NSS and SCA: previous work}
Over the years \NSS has had its fair share of cryptography implementation issues
leading to several practical attacks on multiple primitives---some
of these attacks are algebraic in nature such as
RSA signature forging \cite{DBLP:conf/europki/OiwaKW07},
DH small subgroup attack \cite{DBLP:conf/ndss/ValentaASCFHHH17},
and Lucky13 \cite{DBLP:conf/ccs/ApececheaIES15}, just to name a few.
In 2017, \citet{DBLP:conf/ches/YaromGH16} demonstrated
that cache-bank conflicts leak timing information from an
otherwise constant-time modular exponentiation function implemented in \NSS,
leading to RSA key recovery after observing $16000$ RSA decryptions.
In 2018, \citet{DBLP:conf/sp/RonenGGSWY19} performed a padding oracle attack
against RSA following the PKCS \#1 v1.5 standard to recover long term login
tokens used during TLS connections. Although this attack is well-known,
the authors used recent SCA cache-based attack techniques, successfully
reviving an old vulnerability.
Finally, in 2019 \citet{DBLP:journals/tches/Ryan19} included \NSS in his analysis
of a new SCA attack enabled by a variable-time modular reduction function used
during signature generation, allowing an attacker to recover ECDSA and DSA
private keys.

\subsection{Related Attacks} \label{sec:related_attacks}

\Paragraph{\CVE{2016-2178}} OpenSSL assigned this CVE based on work by
\citet{DBLP:conf/ccs/GarciaBY16}. The authors performed a cache-timing attack
using the \textsc{Flu\-sh+\allowbreak Re\-load} technique against a variable-time
sliding window exponentiation algorithm used during DSA signature generation,
leading to full secret key recovery of OpenSSH and TLS servers co-located with
an attacker. This vulnerability was present in the code base
for more than 10 years and was enabled by a seemingly small software defect.

\Paragraph{\CVE{2018-0737}} OpenSSL assigned this CVE based on work by
\citet{DBLP:journals/tches/AldayaGTB19}. The authors detected and identified several
paths during RSA key generation potentially leaking information about the
algorithm state.
The authors performed a single trace cache-timing attack over the corresponding
GCD function combining different techniques (including lattices) to achieve full
secret key recovery.

\Paragraph{\CVE{2018-5407}} OpenSSL assigned this CVE based on work by
\citet{DBLP:conf/sp/AldayaBHGT19}. The authors discovered a novel
timing SCA attack vector leveraging port contention in shared
execution units on Simultaneous Multi Threading (SMT) architectures.
With a spy process running in parallel, they targeted the variable-time
\wnaf point multiplication algorithm during ECDSA signature
generation and recovered the \ecc{secp384r1} long term private key of a TLS server.
Prior to the CVE and work done by \citet{DBLP:conf/acsac/TuveriHGB18},
this implementation was the default choice for most prime curves,
which was subsequently replaced by a timing resistant version.

\subsection{Leakage Detection and Assessment}

\Paragraph{Differential Address Trace Analysis (DATA)}
This is a framework that detects potential side-channel leaks in program binaries;
\citet{DBLP:conf/uss/WeiserZSMMS18} used the framework to analyze OpenSSL and PyCrypto.
DATA works by observing the program execution with known and different inputs
using Intel Pin\footurl{https://software.intel.com/en-us/articles/pin-a-dynamic-binary-instrumentation-tool},
then analyzing the execution traces to detect differences in flow caused by different
input, thus highlighting potential SCA vulnerabilities.
This approach makes it mostly automated and universal with respect to SCA method.
DATA led to the discovery of \CVE{2018-0734} and \CVE{2018-0735} issued by
OpenSSL \cite{BigNumbersBigTroubles}.

\Paragraph{Triggerflow}
This is a tool to selectively track code-path execution \cite{DBLP:conf/dimva/GridinGTB19},
facilitating testing-based SCA of cryptography libraries such as
OpenSSL and mbedTLS \cite{DBLP:journals/tches/AldayaGTB19, temp:certifiedsca}.
The power of Triggerflow comes from its simplicity, allowing a user to annotate
source code by placing Points of Interest (POIs) and filtering rules, thus
supporting false positive filtering.
Then, Triggerflow compiles the source code, and runs a list of user-supplied
binary invocations called ``triggers'', reporting context whenever a trigger
reaches any of the user-defined POIs.
Triggerflow can be adapted to Continuous Integration (CI) of the development
pipeline for automated regression testing.
Triggerflow does not support automatic POI detection,
instead relying on other offensive methodologies and
tools \cite{DBLP:conf/uss/GrussSM15,DBLP:journals/tissec/DoychevKMR15,
DBLP:conf/uss/WeiserZSMMS18}.
\section{NSS: an SCA Security Assessment} \label{sec:vulns}
In this section, we combine DATA \cite{DBLP:conf/uss/WeiserZSMMS18} for SCA POI
identification with Triggerflow \cite{DBLP:conf/dimva/GridinGTB19} for extended
POI testing. We apply this combination to assess the SCA security of \NSS, in
particular for its public key cryptography primitives.

\Paragraph{DATA frameworks}
DATA requires a ``framework'' for program analy\-sis---a Bash
script defining commands necessary to prepare the environment, run the program
with given inputs and optionally supply a leakage model. The script uses a library
included in DATA and supplies its own domain-specific callbacks. The end result is a script
which accepts parameters such as algorithm, key size and processing phase, and runs DATA.

\Paragraph{An NSS framework for DATA}
For \NSS, we created a framework analyzing signature creation with DSA, ECDSA,
and RSA algorithms.
First, we define DATA callback \code{cb\_\-prepare\_\-framework} which creates
the \NSS certificate storage if it does not exist yet. The storage includes an
SQLite database storing all certificates generated by command-line tools. This
callback runs in the beginning of every framework invocation.
Second, we define \code{cb\_genkey} which generates key pairs and certificates
for a given algorithm using the \NSS utility \code{certutil}. For RSA and DSA we
use default key size, for DSA default parameters, and for ECDSA curves
\ecc{secp256r1}, \ecc{secp384r1}, and \ecc{secp521r1}---the only legacy curves
\NSS features. The callback executes every time DATA needs a different key.
Finally, for DATA analysis we sign a fixed small piece of data using the \NSS
utility \code{pk1sign}
in the callback \code{cb\_\-run\_\-command} traced by DATA.

Supporting DATA code allows us to define all algorithms in a single
file, using algorithm-specific code depending on arguments given in framework invocation.
In our case, the only difference was the algorithm selection during creation of the certificate.
The DATA software package includes example frameworks, as well as working frameworks
for OpenSSL and PyCrypto.

\Paragraph{Performance evaluation}
We performed our experiments on an Intel Xeon Silver 4116 (Skylake) with 256 GB RAM. At peak,
DATA framework consumed 120 GB memory to analyze the program. The exact amount depends
on the stage and algorithm. High memory requirements and general resource consumption
make it unsuitable for automated testing, but still an extremely useful offensive tool for
vulnerability research.

\Paragraph{Results}
DATA output is a collection of potential leaks. It stores the leak data in XML, as well as in Python
standard ``pickle'' serialization format. The included GUI can read this format, and includes tools
for marking leak points for further review, as well as adding comments.
\autoref{tab:data} presents our aggregate DATA statistics, where the ``Total''
rows include also SQLite and/or other less relevant parts of \NSS while
(statically-linked, private) \code{libfreebl} handles the crypto arithmetic in \NSS.

\Paragraph{Combination of DATA and Triggerflow}

DATA can help quickly determine areas of the code vulnerable to SCA,
but it is---as shown before---expensive to run and this is unsuitable for automated testing.
Thus, we combine DATA and Triggerflow in vulnerability research: first, we detect vulnerabilities
once using DATA analysis, then we mark vulnerable areas with Triggerflow annotations and continuously
and cheaply monitor the code for SCA vulnerabilities.
This hybrid approach combines assisted vulnerability scanning of DATA and automatic inexpensive
monitoring by Triggerflow. The approach is general and applicable to any library supported by the tools,
so it can be applied to other cryptographic libraries as well.

\Paragraph{Producing Triggerflow annotations from DATA results}

Using the information from DATA GUI to guide manual code review, we determined
the most critical areas of \NSS potentially vulnerable to SCA. Next, we
annotated each area with Triggerflow's \code{TRIGGERFLOW\_\-POI}, further
refined with \code{TRIGGERFLOW\_\-IGNORE} when running multiple operations on
annotated source code to eliminate false positives. This allows us to examine
potential vulnerabilities in context, and led to several concrete
vulnerabilities summarized in \autoref{tab:summary_attacks} and described in
detail in the following sections.
We further point out that in all the attacks, NSS library was compiled with
debug symbol enabled while keeping the default configurations intact.
\autoref{sec:attack_dsa}--\autoref{sec:attack_rsa} present a more detailed
description of the experiment environments and threat models.

\begin{table}
\centering
\caption{Statistics for our \NSS framework in DATA.}
\label{tab:data}
\begin{tabular}{|l|l|r|r|} \hline
\textbf{Algorithm} & \textbf{Location} & \textbf{CF leaks} & \textbf{Data leaks} \\ \hline \hline
DSA & \code{libfreebl} & 0 & 446 \\
DSA & Total & 2443 & 7435 \\ \hline
ECDSA & \code{libfreebl} & 0 & 1074 \\
ECDSA & Total & 2124 & 5890 \\ \hline
RSA & \code{libfreebl} & 666 & 804 \\
RSA & Total & 3593 & 11140 \\ \hline
\end{tabular}
\end{table}

\Paragraph{Case study}
A good example of this workflow is the vulnerability described later in
\autoref{sec:attack_wnaf}. DATA correctly flagged the problematic line in
\code{ec\_\-compute\_\-wNAF} (\autoref{fig:trace_wnaf_em}), as well as 93 other
potential data leaks.
Inspecting the leak data in DATA GUI showed that all vulnerable places converge
in the parent \code{ec\_\-GFp\_\-pt\_\-mul\_\-jm\_\-wNAF}, which is suitable as
a Triggerflow POI. After running Triggerflow \NSS configuration, \code{pk1sign}
triggered the annotation once for both curves \ecc{secp384r1} and
\ecc{secp521r1}. This annotation could be included in an automatic SCA
regression test for \NSS.

\begin{table}
    \caption{Summary of SCA attacks.}
    \label{tab:summary_attacks}
    \centering
\resizebox{0.47\textwidth}{!}{%
\begin{tabular}{|p{2.5cm}|p{1.4cm}|p{1.5cm}|p{1.8cm}|p{1.5cm}|} \hline
    \textbf{SCA} \newline \textbf{attack}             & \textbf{Vulner\-a\-bility}        & \textbf{Target} \newline \textbf{device}       & \textbf{Application layer}      & \textbf{Threat model} \\\hline \hline             %
    DSA timing \newline (\autoref{sec:attack_dsa})            & Nonce padding        & Raspberry Pi3       & Time Stamp Protocol    & Remote \\ \hline  %
    ECDSA timing \newline (\autoref{sec:attack_padding})          & Nonce padding        & Intel \newline i7-7700       & NSS \code{pk1sign}            & Local \\ \hline   %
    ECDSA Electromagnetic (\autoref{sec:attack_wnaf}) & Point multiplication  & Allwinner Pine A64  & Time Stamp Protocol    & Physical \newline proximity \\ \hline   %
    ECDSA uarch \newline (\autoref{sec:attack_sgx})           & Scalar recoding      & Intel \newline i7-7700 & NSS \code{pk1sign} (SGX)  & Local, malicious OS \\\hline                 %
    RSA Electromagnetic (\autoref{sec:attack_rsa})   & Key generation       & Allwinner Pine A64  & NSS \code{certutil}           & Physical \newline proximity \\ \hline   %
\end{tabular}}
\end{table}
\section{DSA: leakage meets constantness} \label{sec:attack_dsa}
As mentioned previously, after \citep{DBLP:conf/esorics/BrumleyT11}
OpenSSL and several peer vendors decided to apply the nonce fixed bit length countermeasure
to their code base. Unfortunately, this fix did not permeate to the DSA portion
of \NSS, leaving the library vulnerable to this flaw at least since 2011.
We speculate that a very regular fixed window exponentiation (FWE) algorithm
paired with constant-time cache access to pre-computed values provided a false
sense of security, forgetting that the nonce requires its own protection against
bit length leakage due to the fragility of the DSA algorithm w.r.t.\ SCA.

\Paragraph{Analysis}
Our tooling revealed the main root cause of the time
leakage was directly attributed to the variable bit length of the nonce
$k$ during the computation $r = g^k \bmod p$ (\autoref{eq:dsa})
in the upper level \texttt{dsa\_\-Sign\-Digest} function.
Helped by leakage amplification in lower-level exponentiation functions,
this flaw leaks a considerable amount of information on the MSBs of the nonce.
To better understand the time leakage, we can highlight three important functions
in the \NSS library, from general to more specific:
(i) The \texttt{dsa\_\-Sign\-Digest} function contains the logic to calculate the
digital signature pair \textit{(r, s)} by calling the corresponding high-level
modular arithmetic functions;
(ii) The \texttt{mp\_\-exptmod} function is a wrapper function selecting a specific
modular exponentiation function among several available based on input values
and flags set during compilation time---additionally it determines the window size
to be used by the exponentiation function;
(iii) The \texttt{mp\_\-exptmod\_\-safe\_\-i} function computes and implements a
cache-timing safe regular FWE algorithm
based on the window size selected by the previous wrapper function.

After calculating the window size and just before calling the FWE function,
the wrapper function modifies the bit length of the exponent by making it a
multiple of window size. Thus, artificially increasing the amount of bits
in the exponent and therefore the amount of windows to be processed by the
FWE function.
This means the leakage potentially occurs in multiples of the window size, reveling
a total of $w \cdot i$ MSBs for each signature, where $w$ is the window size, and $i$ is
the amount of windows skipped by the FWE due to shorter-than-average nonces.
Therefore, the FWE function effectively amplifies the leakage and improves
the resolution by widening the time gap it takes to process variable length
exponents coming from the upper-level DSA signing function.

\subsection{DSA: Remote Timing Attack}

To demonstrate concretely the impact of the vulnerability, we exploit it
remotely from the application layer through
the Time Stamp Protocol defined in \rfc{3161}
as implemented in
\texttt{uts-ser\-ver}\footurl{https://github.com/kakwa/uts-server}.
The Time Stamp (TS) Protocol permits a trusted Time Stamp Authority (TSA)
to digitally sign a piece of data---\eg using DSA or ECDSA---confirming the data existed at that particular
point in time, and allowing anyone with access to the TSA certificate
to verify the timeliness of the data.

\Paragraph{Target device}
We used a Raspberry Pi 3 Model B plus board containing
a 1.4 GHz 64-bit quad-core Cortex-A53 processor. The device runs stock Gentoo 17,
and we set the board frequency governor to ``powersave''.
We deployed \texttt{uts-ser\-ver} on the target device, acting as the
TSA, receiving TS requests over (the default) HTTP and generating TS responses.

The only supported backend cryptography library for \texttt{uts-ser\-ver} is OpenSSL,
therefore we use an OpenSSL loadable cryptographic
module (engine) \cite{DBLP:conf/secdev/TuveriB19}
to expose \NSS DSA signature generation to the server.
In general, the purpose of engines is to intercept OpenSSL low-level crypto functionality
and carry out the operations internal to the engine, either HW or SW-backed.
The \texttt{e\_nss}\footurl{https://roumenpetrov.info/e_nss} OpenSSL engine
makes the use of \NSS transparent to linking applications, in our case \texttt{uts-ser\-ver}.

As an FOSS contribution, we submitted a PR to the \texttt{uts-ser\-ver} project
adding engine-backed key support, \eg devices such as TPMs, HSMs, or
generically \PKCSONCE driven. In these instances, the key never leaves the
device hence cannot be directly access by OpenSSL, only driven. Our
contribution\footurl{https://github.com/kakwa/uts-server/pull/15}
addresses a three year old outstanding feature request on the
project's issues page. In our case, this allows \texttt{uts-ser\-ver} to
transparently utilize the crypto functionality of \NSS through \texttt{e\_nss}
and keys inside the \NSS keystore through \NSS's \PKCSONCE software token view.

We patched, compiled, and deployed the latest \texttt{uts-ser\-ver} v0.2.0,
linking against an unmodified build of OpenSSL 1.1.1.
We compiled and deployed an unmodified version of \texttt{e\_nss},
linking to both---the previous OpenSSL build and an unmodified build of \NSS \nssVer,
effectively transparently connecting the server to \NSS through OpenSSL.

\Paragraph{Experiment setup}
On one end we deployed the TS server on our RPi, on the other end we deployed
a custom TS client on a workstation equipped with a
3.1 GHz 64-bit Intel i5-2400 CPU (Sandy Bridge), both communicating through a Cisco 9300 series enterprise
switch over Gbit Ethernet. The workstation has 4x1 Gbit link aggregation to the
switch and the RPi a single Gbit connection to the switch.
Our custom client is a simple rust program that embeds a TS request in an
HTTP request, establishing a TCP connection to the server, and starting
a timer just before sending the request. The \texttt{uts-ser\-ver} actively
listens for TS requests and generates corresponding TS responses using \NSS transparently through OpenSSL,
replying back to the client as soon as the TS response is available.
Once the TS response is received, the client stops the timing, closes the TCP
connection, computes the latency, and finally stores the latency and the TS
response pairs in a database. This operation is repeated as needed to gather more samples.
Following the attack methodology from previous remote
attacks \cite{DBLP:conf/esorics/BrumleyT11,temp:tmpfail,temp:certifiedsca},
we divided our attack in two phases.

\begin{figure}
\iftrue
\includegraphics[width=\linewidth]{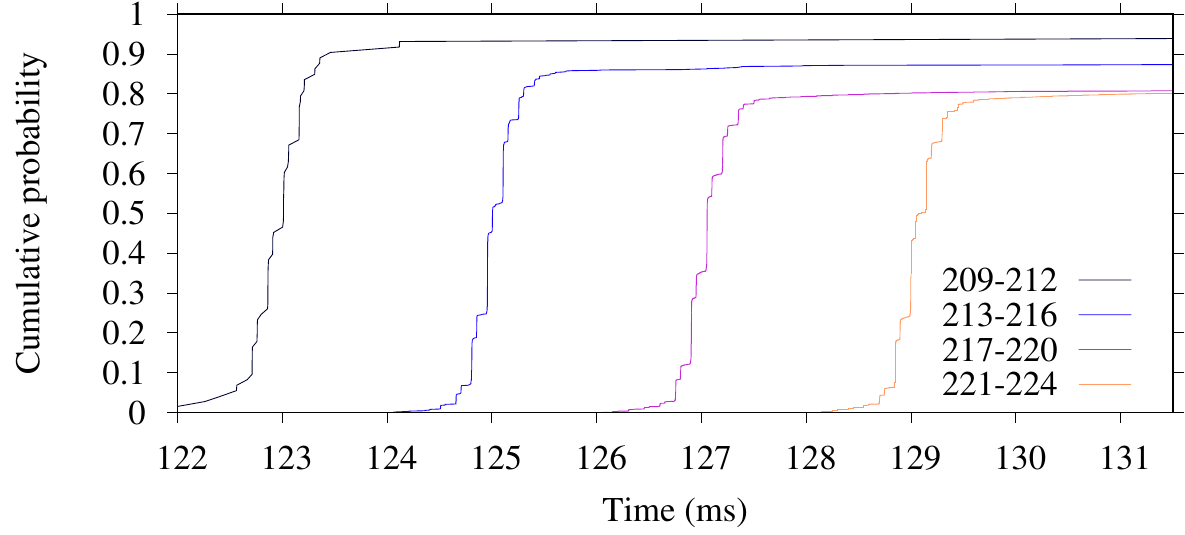}
\else
\Huge\bf PLACE\\HOLDER
\fi
\caption{DSA time leakage in \NSS on a remote scenario. Direct correlation
    between DSA signature generation time and bit length of the nonce.}
\label{fig:dsa_remote}
\end{figure}

\Paragraph{Collection phase}
We collected $2^{18}$ samples using our custom client,
and our timing analysis confirmed that our vulnerability analysis was
correct---a direct correlation existed between the wall clock execution time of
DSA signature generation and the bit length of the nonce used to compute the
signature. We confirmed the nonce values using the ground truth private key
from the \NSS keystore.
More importantly, \autoref{fig:dsa_remote} shows that the time leakage is
substantial, allowing an attacker to exploit the vulnerability even in a
remote scenario.

\Paragraph{Recovery phase and results}
\autoref{sec:lattice} describes in detail the lattice construction
formalization, lattice parameters, lattice experiments, and results applied
to our collected samples.
In short, we observed a striking 99\% (1536 samples) and 38\% (1152 samples)
success rate for recovering private keys in our remote timing attack scenario
after performing a lattice attack.

\section{Are your nonces really padded?}%
\label{sec:attack_padding}

In a nutshell, \cveRemoteStill exploited the Montgomery ladder feature that it
executes an iteration per scalar (\ie ECDSA nonce $k$) bit, performing both \dbl
and \add ECC operations regardless of said nonce bit value. This regularity offers
protection against classical SCA that aims at recovering the nonce by tracking
the sequence of ECC operations \citep{DBLP:conf/crypto/KocherJJ99}. However as
demonstrated in \citep{DBLP:conf/esorics/BrumleyT11} this feature plays in favor
of a timing attacker.

This highly regular feature combined with the fact that this algorithm executes
$\ceil{\lg k} -1$ iterations implies that the execution time is highly related
to the effective bit length of $k$
\citep{DBLP:conf/esorics/BrumleyT11,DBLP:conf/acsac/TuveriHGB18}. Therefore, a
timing attacker could learn information on $k$ by measuring the execution time
during ECDSA signature generation, with computing time dominated by the scalar
multiplication.

In order to prevent this attack, \citet{DBLP:conf/esorics/BrumleyT11} proposed a countermeasure
that fixes the number of significant bits of the nonce using \eqref{eq:nonce_padding}.
\begin{equation}
\label{eq:nonce_padding}
\hat{k} =
    \begin{cases}
        k + 2q &\textnormal{if} \,\,\, \ceil{\lg (k + q)} = \ceil{\lg (q)} \\
        k + q  &\textnormal{otherwise}
    \end{cases}
\end{equation}
In response to this research and \cveRemoteStill,
the \emph{nonce padding} countermeasure based on \eqref{eq:nonce_padding} has been
implemented not only in OpenSSL, but in other libraries as well.
Mozilla \NSS included the nonce padding countermeasure in 2011 in their high-level ECDSA function%
\footurl{https://hg.mozilla.org/projects/nss/rev/079cfc4710c7193ef73888394f4d4f935e03f241}.
However, despite this intended fix, we uncovered the implemented countermeasure is ineffective.
We found that after the padding,
a lower-level scalar multiplication function reduces $\hat{k}$ modulo $q$,
thus reverting the nonce to its original value
(Line 25\footurl{https://hg.mozilla.org/projects/nss/file/c06f22733446c6fb55362b9707fa714c15caf04e/lib/freebl/ecl/ecl_mult.c}).

This \emph{nonce unpadding} opens the door to a timing attack against \NSS.
The library has different scalar multiplication algorithms implemented. For
curve \ecc{secp256r1} it uses a constant-time scalar multiplication algorithm,
yet a \wnaf implementation for higher security curves \ecc{secp3841} and
\ecc{secp521r1}. This algorithm iterates through \wnaf digits of the scalar, and
the number of iterations depends on the \wnaf representation length, eventually
depending on $\lg(k)$. Therefore a timing attacker could learn information about
$k$ by measuring the duration of ECDSA signature generation.

\Paragraph{Experimental validation}
In order to validate this hypothesis, we developed a proof-of-concept to
demonstrate that the execution time of \NSS \wnaf scalar multiplication is
related to $\lg(k)$. For this experiment we build NSS \nssVer on an Ubuntu 18.04
LTS desktop workstation running on Intel i7-7700 3.60GHz (Kaby Lake). We
measured the number of clock cycles consumed by the NSS library exported
function for generating digital signatures: \code{Sign\-Data}. Our previous
analysis (\autoref{sec:vulns}) reveals this function's call trace includes
\code{ec\_\-GFp\_\-pt\_\-mul\_\-jm\_\-wNAF} for $w=5$ in the context of
\ecc{secp384r1}.
We collected 1M samples of \code{Sign\-Data} latency during ECDSA signature
generation. Using the ground truth private key from the NSS keystore, we
computed the secret nonces used in each signature, then estimated the cdf curves
of the latency per effective bit length.

\begin{figure}
\iftrue
\includegraphics[width=\columnwidth]{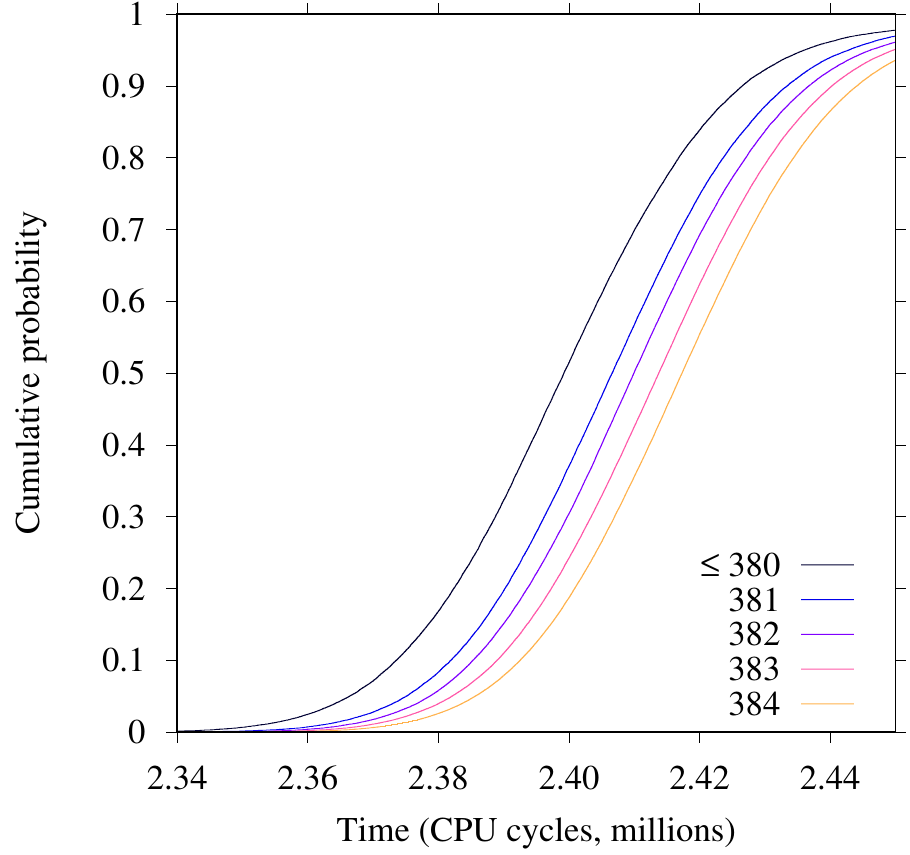}
\else
\Huge\bf PLACE\\HOLDER
\fi
\caption{ECDSA timing cdf per nonce bit-length.}\label{fig:ecdsa_timing}
\end{figure}

\autoref{fig:ecdsa_timing} shows these curves, aggregating those $\ell \leq 380$
in one single curve. This empirically demonstrates there is indeed a dependency
between $\lg(k)$ and the time taken to produce an \ecc{secp384r1} ECDSA
signature in NSS. With enough samples under the right conditions, we speculate
this leak could be exploited using the lattice methods developed in \autoref{sec:lattice},
as demonstrated in \cite{temp:certifiedsca}.

\section{Leaking ECDSA keys through EMA} \label{sec:attack_wnaf}

In general, \wnaf has been subjected to a variety of SCA attacks on OpenSSL in
the past---\eg L1 and LLC cache timings
\citep{DBLP:conf/asiacrypt/BrumleyH09,DBLP:conf/ches/BengerPSY14}, EM
\citep{DBLP:conf/ccs/GenkinPPTY16,temp:certifiedsca} and port contention
\cite{DBLP:conf/sp/AldayaBHGT19}. As far as we know, we are the first to
practically demonstrate an end-to-end attack on NSS \wnaf implementation. To
this end, we employed EMA to exfiltrate the ECDSA
private key. Previous EM attacks on \wnaf focused on only retrieving
least significant bit
positions \cite{DBLP:conf/ctrsa/BelgarricFMT16,DBLP:conf/ccs/GenkinPPTY16,temp:certifiedsca}.
In contrast, out attack uses an advanced multi-digit lattice formulation detailed
in \autoref{sec:lattice_wnaf}, making it possible to potentially use the
entire EM trace to extract side-channel information, consequently,
lowering the number of signatures required and reducing the data
complexity of the attack.

\Paragraph{Threat model}
We assume an adversary is able to obtain a
similar device to learn about the particular EM leakage (preparation phase),
and furthermore gain access to close proximity of the
target device while issuing ECDSA queries (attack phase).
This model is consistent with the literature.

\Paragraph{Experiment setup}
Our setup includes a Pine A64-LTS powered by a 64-bit quad-core ARM Cortex A53
SoC. This target hardware device runs Ubuntu $16.04.5$ LTS minimal with NSS
\nssVer. Similar to \autoref{sec:attack_dsa}, we created a TS server instance
using \texttt{uts-ser\-ver} as our victim, this time with an \ecc{secp384r1} key.
We measured the EM signals using a Langer LF-U $2.5$ EM probe
attached to a $40$ db preamplifier, with the probe head positioned
close to the target board to achieve good signal strength.
We acquired the EM traces using a Picoscope $6404C$ USB digital oscilloscope
with a maximum sampling rate of $5$GSps supporting up to $500$MHz bandwidth.
To strike balance between lower computational cost and decent
signal-to-noise ratio, we used a sampling rate of $150$MSps instead.

\Paragraph{Signal acquisition}
We created a client responsible for sending TS requests over HTTP to our
server and controlling the oscilloscope. The client first initiates the trace
capture command followed by a TS request, then stopping the trace capture
upon receiving the HTTP response from the server. We parsed the server response
messages to retrieve DER encoded ECDSA signatures and the hash from the client
request. The resulting EM traces along with their parsed ECDSA information were
stored for offline signal processing and key recovery phase.

\begin{figure}
\iftrue
\includegraphics[width= 1\linewidth]{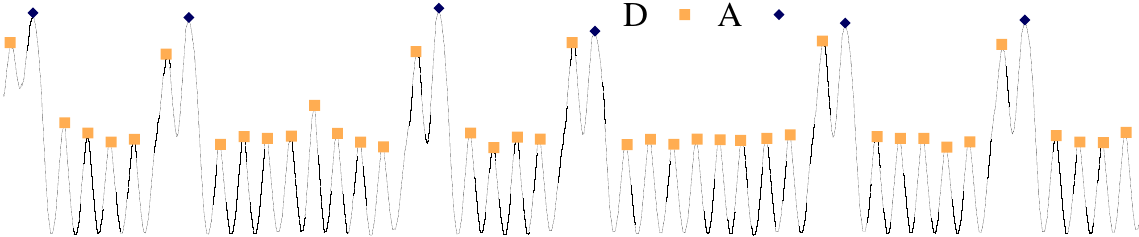}
\else
\Huge\bf PLACE\\HOLDER
\fi
\caption{EMA trace showing part of the \wnaf point multiplication with
marked \dbl (D) and \add (A) operations. The filtered trace clearly shows two distinguishable
loops: lower energy D and higher energy DA.
}
\label{fig:trace_wnaf_em}
\end{figure}

\Paragraph{Signal processing}
For a successful signing key recovery, the EM traces must go through signal processing
to reliably extract the partial nonce information. From the signal analysis
perspective, these partial nonces are encoded as the sequences of \dbl and \add operations
during \wnaf point multiplication as observed in the EM trace (\autoref{fig:trace_wnaf_em}).
Using this partial information from multiple signatures we can formulate a
lattice attack as described in \autoref{sec:lattice_wnaf}.

Since the captured trace contained the entire TS request window, the first step involved
in locating and isolating the ECDSA point multiplication part. By performing a manual
analysis, we found specific patterns in the trace pertaining to the start and
end of the point multiplication. We used these as the templates (created by averaging
over $20$ EM traces) to cut the point multiplication window using squared
Euclidean distances between root mean square values of the trace and template.

We then moved to the next phase, extracting the \dbl and \add sequences.
This was a two-step process: finding the position of all the \add operations
and then finding all {\dbl} operations between them. By performing a spectrum analysis we found
clearly distinguishable low energy point multiplication \dbl loop ($D$) and
higher energy \dbl and \add loop ($DA$). To improve the detection
we extracted two components of the signal: a band pass around 15 MHZ and
a low pass at 5Mhz for the $DA$ and $D$ loops respectively. We
demodulated them using a digital Hilbert transform and applied
signal smoothing filter.

Using the first signal component, we extracted the $DA$ loops using a similar
approach to the point multiplication extraction, \ie compute rolling squared Euclidean
distances using the $DA$ template. The $D$ sequences were present in the signal as voltage
peaks, however due to noisy artifacts in the trace simply extracting the peaks resulted in
both false positive and negative peaks. Since each \wnaf loop iteration performs only one
\dbl operation between two $DA$ loops, they follow an almost periodic trend.
Using this information together with the fact that $D$ peaks can be approximated to
rectangular pulses using root mean square, we computed pulse width to period ratios.
By selecting an experimentally evaluated threshold for these ratios we were able to
significantly increase the detection of the $D$ loops with an overall error rate of less then $1\%$.

In practice, EM traces contain noise from various sources---OS preemption,
acquisition noise, environmental and electrical noise---which can reduce the efficacy
of signal processing phase. OS interrupts for instance are high energy signals and
therefore easily distinguishable. We marked all such interrupts in the trace and
recovered the sequence from the interrupt position till the end of the trace
(\ie interrupt to \wnaf LSD). For other noise sources and small interrupts our
sequence extraction resulted in around $6\%$ of the total traces with less then $1\%$
incorrect guess for $D$ and/or $AD$ loops. Additionally, we applied heuristics on
the recovered sequences to filter those which violated the \wnaf encoding rules.

In total we collected $300$ signatures, which left us with $211$ signatures
after performing sequence extraction, containing $13$ errors. We filtered out the sequences
with a length of at least $384$, which resulted in a total of $66$ signatures.
By using our lattice formulation, we were able to recover the private key
with as few as $30$ signatures as described \autoref{sec:lattice_wnaf}.
A clear advantage of using more information per trace is reflected in the low
number of signatures required. To put this into perspective, the ECDSA attack
presented by \citep{DBLP:conf/ccs/GenkinPPTY16} utilized only LSDs of the
nonce (last non-zero digit and trailing zeros), required $3060$ signatures and
even the \ecc{secp256k1} curve at a substantially lower security level.

\section{Uarch SCA on Scalar Recoding}\label{sec:attack_sgx}

This section presents an SCA attack that aims at recovering the \wnaf
representation of an ECDSA nonce, similar to \autoref{sec:attack_wnaf} that
targeted the ECC operation sequence.
We instead target the \wnaf recoding itself (\autoref{alg:compute_wanf}).
This target, in addition to representing a novelty as it is not a common target in the literature,
represents a challenge due to its low temporal and spatial granularity as detailed later.
Moreover, we will be able to recover the same information as in \autoref{sec:attack_wnaf}.
Additionally, employing the same channel we will gain extra information on the \wnaf representation,
recovering the sign of its non-zero coefficients.

\Paragraph{Threat model}
Although \autoref{sec:attack_wnaf} and this section aim at recovering very related data,
their threat models and targets differ significantly, therefore they are distinct attacks.
Controlled channel attacks belong to a class of threat models when the adversary
has control over the targeted computing platform except the targeted algorithm itself and its processed secrets
\citep{DBLP:conf/sp/XuCP15}.
One instance of this threat model is provided by trusted executed environments (TEEs),
like Intel Software Guard Extensions (SGX) on Intel microprocessors \citep{DBLP:journals/iacr/CostanD16}.
In Intel SGX nomenclature, an \emph{enclave} is software running on a secure space
that provides confidentiality and integrity of \emph{enclave} code and data,
even in presence of a compromised OS (\ie adversary).
However, it delegates SCA protections to developers, allowing attackers to use
privileged OS resources when gathering SCA signals that reveal information on
enclave secrets.

Intel SGX leaves control of memory pages to the OS, which an adversary can use
to track the sequence of executed memory pages by a targeted
enclave \citep{DBLP:conf/sp/XuCP15,DBLP:conf/ccs/WangCPZWBTG17,DBLP:conf/ccs/ShindeCNS16,DBLP:conf/uss/BulckWKPS17,DBLP:conf/ccs/WeiserSB18}.
An attacker first marks a memory page with SCA relevance as
\emph{non-executable} and launches the enclave. If the enclave executes that memory page,
a page fault is generated and handled by the OS (\ie adversary), hence
the attacker learns the targeted memory page was executed \citep{DBLP:conf/sp/XuCP15}.
Applying this process for a set of memory pages allows the adversary to track the sequence
of executed memory pages, thus potentially leaking secret data processed by the enclave.
This attack works at 4KB granularity, yet sufficient to recover some secrets
on low granularity targets as detailed below.

\Paragraph{Experiment setup}
To track the sequence of executed memory pages of an enclave, we used
the \sgxstep framework proposed by \citet{DBLP:conf/sosp/BulckPS17}
and integrated into the \graphene framework \citep{DBLP:conf/usenix/TsaiPV17}.
\graphene allows running unmodified code inside an Intel SGX enclave,
providing a straightforward approach to execute \NSS code in an enclave
and assess its SCA resistance.
It is worth noting the \graphene framework is not a requirement
for the attack, it just simplifies porting \NSS to SGX.
We performed our experiments on Ubuntu 18.04 LTS running on a desktop workstation
featuring an Intel i7-7700 microprocessor (Kaby Lake) with Intel SGX enabled.

\begin{figure}[h]
\centering
\iftrue
\includegraphics[width=0.6\columnwidth]{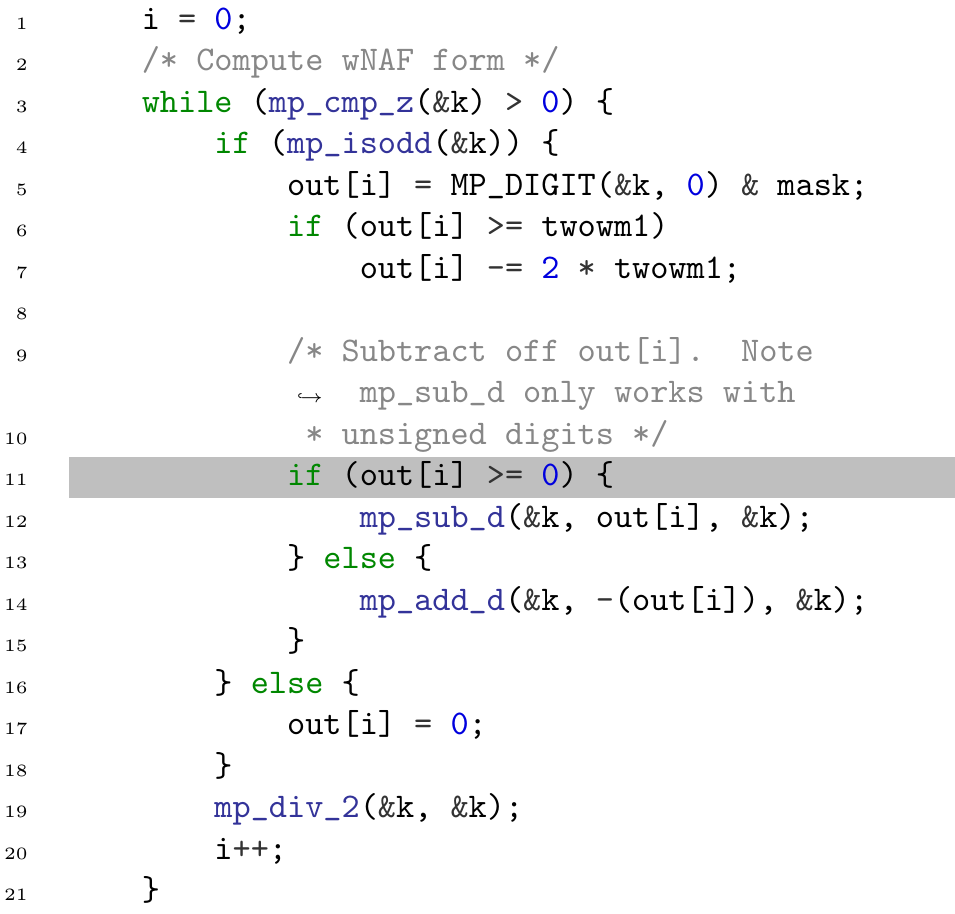}
\else
\Huge\bf PLACE\\HOLDER
\fi
\caption{NSS \wnaf encoding function snippet.}
\label{fig:sgx_target}
\end{figure}

The \NSS (private) function \Libwnaf computes the \wnaf encoding.
\autoref{fig:sgx_target} shows a snippet of this function,
consisting of a main loop that encodes $k$ into its \wnaf representation, which is stored in \code{out}.
In our build this snippet compiles to 363 bytes, hence much smaller than a memory page;
however its callees are located on different pages.

\newcommand{\mintedLine}[1]{#1\xspace}
\newcommand{\lineLoop}{\mintedLine{3}}
\newcommand{\lineOdd}{\mintedLine{4}}
\newcommand{\lineSign}{\mintedLine{11}}

The execution flow of this function is related to the \wnaf representation of $k$ to different degrees.
For instance, it is easy to verify the results of conditions at lines \lineLoop and \lineOdd allows
retrieving the indices of the non-zero coefficients of the \wnaf representation.
A further analysis revealed it is also possible to extract the sign of these coefficients,
by inferring the condition result at line \lineSign.
This additional information will reduce lattice computation time to recover
ECDSA keys as shown in \autoref{sec:lattice}.
Note that this \emph{sign leakage} is due to \NSS API \LibSub only supporting unsigned digits as commented in \autoref{fig:sgx_target}.
This is just another example that the implementation has the final word
regarding SCA (\cf \autoref{alg:compute_wanf}).

\autoref{tab:pages} shows the relation between functions of interest for SCA and
the tracked memory pages we used to record the execution flow of \Libwnaf.
The first three memory pages allow extracting the unsigned non-zero coefficients of the \wnaf representation,
while additionally tracking \LibCmpDig allows sign recovery.

\begin{table}
\centering
\caption{Functions of interest and tracked pages.}
\label{tab:pages}
\begin{tabular}{|l|c|c|}
    \hline
    \textbf{Function} & \textbf{Memory page} \\
    \hline
    \hline
    \Libwnaf         & \code{0x08000} \\ \hline
    \LibIsOdd        & \code{0x1f000} \\ \hline
    \LibAdd, \LibSub & \code{0x22000} \\ \hline
    \LibCmpDig       & \code{0x24000} \\ \hline
    \end{tabular}
\end{table}

The memory page sequence of \LibAdd and \LibSub are almost
identical. However, a subtle difference allows distinguishing them. \LibSub call
trace reveals \LibCmpDig executes more times in \LibSub than in \LibAdd,
making it a good \LibCmpDig distinguisher to determine the condition result at
line \lineSign in \autoref{fig:sgx_target}.

We developed an SGX enclave that generates ECDSA signatures using curve
\ecc{secp384r1} through \NSS \code{pk1sign}. We targeted this enclave collecting 1000 traces
while tracking the memory pages in \autoref{tab:pages}. Using the ground truth
private key, we verified the non-zero coefficient signs of the nonce \wnaf
representation used to generate those signatures were perfectly recovered in all
cases.
As detailed in \autoref{sec:lattice_wnaf}, after applying lattice cryptanalysis
we were able to recover the private key with very high probability.
\section{Leaking RSA keys through EMA} \label{sec:attack_rsa}

We now present another EM attack: exploiting the BEEA algorithm during RSA key generation.
During NSS RSA key generation,
the function \code{RSA\_\-PrivateKeyCheck} makes two calls to the vulnerable
function \code{mp\_gcd} to validate if the public exponent $e$ is relatively
prime to $p-1$ and $q-1$.

\Paragraph{Threat model}
Our attack utilizes a single EM trace to recover the private key, since the
attacker only gets one shot at the key. Hence our model assumes the attacker can
either trigger RSA key generation or knows when it occurs. The threat model is
otherwise the same as in \autoref{sec:attack_wnaf}.

\Paragraph{Experiment setup}
For capturing EM traces during RSA key generation,
we use the same setup and target device described in \autoref{sec:attack_wnaf}.

\subsection{Signal Acquisition and Processing}\label{sec:rsa_signal}
Using the NSS \code{certutil} tool, we issued
self signed certificates requesting a fresh $2048$-bit RSA key pair each time,
while ensuring sync with the oscilloscope's signal capture window.
We logged the key metadata and the corresponding EM traces for further analysis
and key recovery.
We captured $1100$ independent traces and used $100$
as a \emph{training set} to adjust signal processing and error correction
phases of the attack.
The remaining $1000$ are left to present statistics of the proposed attack (\autoref{sec:rsa_results}).

To increase the success rate of key recovery, we preprocessed the traces to
remove high frequency noise and detect noise sources such as interrupts. We
applied low pass filter followed by a digital Hilbert transform. We then moved
on to extracting $p$ and $q$ traces from the entire key generation trace. By applying
templates obtained from two distinctive patterns, at the start of the first
and the end of the second GCD computation, we extracted the signal of interest.
Finally to cut the traces into $p$ and $q$, we used the fact that the
signal mean changes abruptly when the second GCD computation starts, creating
a distinctive dip in the trace.

\begin{figure}
\iftrue
\includegraphics[width= 1\linewidth]{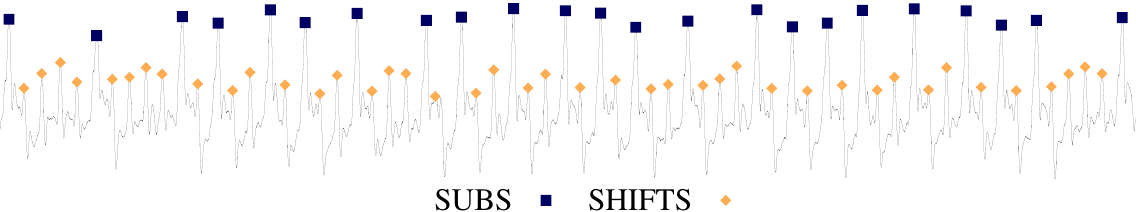}
\else
\Huge\bf PLACE\\HOLDER
\fi
\caption{Window selected from a random trace. SUBS operations represented by
peaks and SHIFTS by the distance among them.}
\label{fig:trace_rsa}
\end{figure}

After the trace preprocessing, we selected a single trace to serve as a
template. We divided this template trace into 17 windows (\cf
\autoref{fig:trace_rsa}). For each window we used peak extraction to identify
SUBS operations along the trace, while the distance between those peaks relates
to the number of SHIFTS operations between two SUBS. The distance between peaks
is decreasing along the trace, as operations require less and less time as
the integers $u$ and $v$ from \autoref{alg:BEEA_alg} decrease in magnitude.  For
each window we create a linear regression model that relates the distance
between peaks with the number of SHIFTS operations produced between two SUBS
operations. This way, the unit distance for a SHIFT operation is not the same
for different windows, and we empirically found 17 to be the minimum number of
windows containing the maximum possible number of peaks, for linear regression
models to accurately relate distances with SHIFT operations. Once the 17 models
are generated, it is possible to select any trace, cut it in 17 windows with the
same number of peaks within them and recover the entire sequence of operations
made by the BEEA, applying the regression model corresponding to each window.

In a noise-free scenario, using the whole sequence of SHIFTS and SUBS
it is possible to recover the primes \citep{DBLP:journals/ijcta/AldayaMSS17,DBLP:journals/tches/AldayaGTB19}.
However, due to the noisy nature of EMA,
errors possibly exist in the recovered sequences for $p$ and $q$.
Therefore, instead of trying to recover the full sequence we aimed to
recover a partial one with sufficient information to retrieve a prime.
\citet{DBLP:conf/eurocrypt/Coppersmith96} proposed an algebraic approach
that allows to recover about half the bits of a prime knowing the other half
employing lattice methods.
Using \cite{DBLP:conf/eurocrypt/Coppersmith96}, \citet{DBLP:journals/tches/AldayaGTB19}
showed it was practically possible to recover a 1024-bit prime knowing its 522 least significant bits
with very high probability in less than 10 minutes.
Therefore, we adopted a similar approach to \citep{DBLP:journals/tches/AldayaGTB19}, taking into
account its lattice implementation is open source.
In what follows we treat this lattice-based cryptanalysis
as a \emph{lattice oracle} that takes 522 bits of a prime as input and outputs the remaining bits.
Following established leakage models of the BEEA \citep{DBLP:journals/jce/AldayaSS17,DBLP:journals/tches/AldayaGTB19},
for recovering $t$ bits we need to recover a sequence
such that the number of total SHIFTS is at least $t$.
This implies we need to gather about half a trace (\ie $t=522$),
considerably reducing the influence of noise.

\subsection{Error Correction of Noisy RSA Keys}\label{sec:rsa_general_error_correction}
In each key generation, the function \texttt{mp\_gcd} executes for both $(e, p-1)$
and $(e, q-1)$, thus we have two traces per key generated. Hence two shots to
recover one of the secret primes, $p$ or $q$, needed to compute the private key.
We address the problem of retrieving the bits from the recovered sequence using
two different approaches.
The first approach consists of trying to retrieve the prime using only the
recovered sequence from the trace corresponding to a single prime, with
a classical brute force mechanism.

Our linear regression
models can make different mistakes during the task of retrieving the SHIFTS count
between SUBS peaks, and a single error could lead to incorrect recovery
of the secret prime. For that reason, it is necessary to detect components that
give hints indicating error due to the way they were determined.

\Paragraph{Error modeling}
During our sequence recovery process we noticed there are three
frequent error types:
(i) The model finds a distance corresponding
to 0 SHIFTS between two SUBS operations, which is a contradiction; thus, we
delete that component from the vector, and consider the surrounding components
susceptible to error. This normally happens because the acquisition of
the peak is not clean, and we find in the trace two peaks extremely near to one
another, that actually should be represented by only one peak.
(ii) The model recovers the number of SHIFT operations from linear regression
and requires rounding. As the fractional part nears $0.5$ the model can decide
incorrectly---we consider every component with its fractional part in
the range $[0.3, 0.5]$ to be error-susceptible.
Finally, (iii) we noticed that for
high SHIFTS count the regression model accuracy decreases. This is due to the
fact that a higher number of SHIFTS is increasingly improbable, so we had
considerably less samples of high SHIFTS count during the
procedure to generate our linear regression models. For this reason, we consider
every component finding a SHIFTS count exceeding 8 to be error-susceptible.

To fix these potential errors, we propose the following corrections:
(i) Modify the component by $\pm 1$, allowing
three possibilities for each component (counting the original retrieved);
(ii) Modify the component by $\pm 1$ depending if the rounding operation leads
to the next or previous natural number (two possibilities);
(iii) Modify the component by $+1$ and $+2$ (three possibilities) since, from an
analysis done in the training set, we found the model tends to underestimate the
real value when the SHIFT count between two peaks is relatively high.

Finally, it is important to note that our model is not able to retrieve the first iterations
of the algorithm, \ie the SHIFTS and SUBS operations are unknown
during these iterations. However, we observed that no more than four SUBS operations
were lost at the beginning of the sequence. For that reason, we consider all
possible combinations of lost SUBS operations up to four
and up to six SHIFTS\footnote{Higher values are possible, but increasingly improbable.}
between SUBS, leading to a total of $\sum_{z=0}^46^z = 1555$ possibilities to recover
the leading sequence of SHIFTS and SUBS operations.

\Paragraph{Feasible exhaustive search error correction}
Concerning the brute force procedure, we used the training set
to recover statistics about the worst case computational cost of this approach.
Note in this approach the attacker has two shots per trace to recover the first 522 bits of a prime.
After our analysis, we verified that all traces suffer from missing iterations at the beginning,
therefore the attacker needs 1555 calls to the lattice oracle to succeed in the worst case.
At the same time, we successfully retrieved 14 out of 100 keys with this low computational
cost---the recovered trace matched the ground truth sans the missing iterations.
Generally, considering a \emph{low} computational cost adversary (\ie less than 17 error hints)
it is feasible to recover 51 out of 100 private keys using only the bruteforce approach.
In the first row, \autoref{tab:rsa_stats} gives the statistics about the associated worst case computational costs.
However, beyond these numbers, we highlight it is feasible for an adversary to
solve this problem using an exhaustive search approach.

\Paragraph{Combined error correction}
It is also possible to exploit the leak redundancy in the EM traces
of $p$ and $q$, combining them to fix errors.
This is a common technique to recover noisy RSA keys,
useful in our scenario considering the noisy nature of an EM side-channel attack.
It was first introduced by \citet{Percival05} and later extended and formalized by \citet{DBLP:conf/crypto/HeningerS09}.
For surveys about error correction in noisy RSA keys the reader can consult
\citep{DBLP:conf/crypto/HeningerS09,MR3586868,DBLP:journals/tches/AldayaGTB19}.

In our work, we use the approach proposed in \citep{DBLP:journals/tches/AldayaGTB19}
due to the similarity between the error types handled there and our observations.
Their error correction approach belongs to the binary \emph{extend-and-prune} algorithm class.
Where, in our RSA context we obtain a binary sequence that represents the execution flow
as explained in \autoref{sec:rsa_signal}.
However, despite reusing their algorithm
we use it in a different scenario not originally considered,
which we expand on during the experimental evaluation.

A binary \emph{extend-and-prune} algorithm for fixing errors in RSA keys
processes noisy binary sequences \emph{expanding} them to a set of candidates considering possible error sources:
(i) error at $p_i$, (ii) error at $q_i$, (iii) error at both $p_i$ and $q_i$, (iv) no error at all.
This expand process ensures that the relation $N=pq \bmod 2^i$ holds, where $i$ represents the sequence index.
Avoiding candidate space explosion requires a \emph{prune} procedure.

The \emph{prune} approach proposed in \citep{DBLP:journals/tches/AldayaGTB19}
is to use a set of filters over the candidates, \eg
hard limiting the number of candidates that will not be pruned,
total number of errors since start (\ie promoting those candidates with less errors since start).
For a full description on the filters employed we refer the reader to \citep{DBLP:journals/tches/AldayaGTB19}.
In our work we used an unmodified version of the error correction algorithm employed in \citep{DBLP:journals/tches/AldayaGTB19}.

This algorithm could output a maximum of 150k candidates per trace,
indicating that a ranking would be useful to reduce the number of calls to the lattice oracle.
This algorithm incorporates a candidate enumeration approach
that, according to the experiments in \citep{DBLP:journals/tches/AldayaGTB19},
ranks the correct solution at first position with high probability.

\subsection{Extended Experiment Results}\label{sec:rsa_results}
For this extended experiment we used $1000$ EM traces as described in \autoref{sec:rsa_signal}.
From each of these traces we extracted the sequences of SHIFTS and SUBS corresponding to the processing
of $p$ and $q$.
We used the training set to adjust the combined error correction algorithm,
comparing observed trace recovery with the ground truth.
Hence we decided to use the same configuration employed in \citep{DBLP:journals/tches/AldayaGTB19}.

As described during the exhaustive search approach for solving errors in $p$ or $q$,
it is common that traces lack information about the first iterations,
also in some of them recovering an additional iteration.
These errors cannot be handled by this error correction algorithm,
therefore implying a failed recovery \citep{DBLP:journals/tches/AldayaGTB19}.

In our training set, the number of missing iterations ranges from zero to four.
A straightforward approach to solve this was explained in \autoref{sec:rsa_general_error_correction},
generating 1555 candidates per prime trace.
Therefore, considering the traces of $p$ and $q$ the number of candidates
exploded to $1555^2 \approx 2^{21}$.
It is then possible to launch the error correction algorithm and see which one gives a solution.
However, while this method is feasible,
we decided to explore another approach that considerably reduces the number of candidates.

Instead of exhaustive searching the number of missing leading iterations and
the SHIFTS count inside them,
we bruteforce only the iteration count and fix SHIFTS in all missing iterations to one.
It is likely these filled iterations contain errors,
but our hypothesis is the error correction algorithm will be able to fix them automatically.
With this approach the number of candidates reduces from $2^{21}$ to 25.

Due to this considerable reduction, in addition to attempting to solve missing iterations
we decided to handling cases where traces have an additional leading iteration.
Removing the first iteration and treating it as a missing one gives a total of 36 traces
per original trace.
This approach was not considered in \citep{DBLP:journals/tches/AldayaGTB19},
where the authors confirmed that 30\% of their traces have missing information at the beginning,
but did not solve it---a gap our work fills.

We tested this approach on the training set, obtaining a success rate of 64\%.
After manually inspecting the training set, we observed a maximum success rate
of $67/100$ using this combined error correction algorithm. In that sense,
$64/67$ is sufficiently high for our purposes.

One interesting feature is the algorithm spends considerably less time
processing a trace with the correct number of iterations at the beginning
compared to incorrect. This is important because the computing time is one good
indicator if there is a solution in the processed trace. This allows us to
quickly detect which trace out of the 36 has the correct fix for the missing
iterations.

After this training, we applied this procedure with the same configuration
parameters in a large set to estimate the success rate of the full attack.
Fixing the number of bits to recover to 522, we expanded 1k traces on their 36
candidates each and executed the error correction
algorithm on all of them (36k) limiting the computing time to 15 min per trace.
The results are as follows: 587 traces terminated in time and after recovering the remaining
bits using well-known lattice techniques, we recovered 565 independent RSA-2048 private keys.
The number of calls to the lattice oracle had an impressive median of one while achieving a success rate of 56.5\%.
\autoref{tab:rsa_stats} shows more statistics about this method. 

\begin{table}
\centering
\caption{Number of calls to the lattice oracle.}
\label{tab:rsa_stats}
\renewcommand{\arraystretch}{1.5}
\begin{tabular}{|l|c|c|c|c|}
    \hline
    \textbf{Method} & \textbf{Min} & \textbf{Median} & \textbf{Mean} & \textbf{Max} \\
    \hline
    \hline
    Bruteforce      &    $1555$      & $2^{22}$ & $2^{26}$ & $2^{30}$ \\ \hline
    Combined        &     $1$        &    $1$   &   $3$    & $720$    \\ \hline
\end{tabular}
\end{table}

Data show in \autoref{tab:rsa_stats} demonstrates that an EM attack
on binary GCD algorithms is a real threat to SoC devices.
This table is not about comparing two approaches: rather providing experimental
data of attack feasibility using different methods, where both have room for improvements.

\section{Endgame: lattice-based analysis} \label{sec:lattice}

{ECDSA} and {DSA} signing both return a pair \((r,s)\) such that
\begin{align}
        s\cdot k \equiv h + \alpha \cdot r \bmod q
        \label{eq:rs-sign}
\end{align}
where \(\alpha\), \(h\), and \(k\) correspond with the private key, hash, and nonce.
Additionally, the integer \(r\) coincides with the output of a procedure that
performs an exponentiation ({DSA}, \autoref{eq:dsa}) or point multiplication
({ECDSA}, \autoref{eq:ecdsa}).
The vulnerabilities in \autoref{sec:attack_dsa}--\autoref{sec:attack_sgx}
provide varying degrees of leakage for each nonce, and in this section we
utilize that for lattice-based cryptanalysis.

The private key recovery problem reduces to a Shortest Vector Problem (SVP) or
Closest Vector Problem (CVP) instance of a given lattice. In both SVP and CVP
instances, one proceeds by reducing the lattice with the LLL or BKZ procedures;
the next step is looking for a short or close vector on the lattice,
respectively. However, any CVP instance with input lattice \(B\) and vector
\(u\) can be mapped into an SVP instance by looking for a short lattice basis
vector in the lattice
\begin{equation}
        \hat{B} = \left[
        \begin{array}{c|c}
                B       & \vec{0} \\
                \hline
                \vec{u} & q
        \end{array}\right]
        \label{eq:lattice:svp}
\end{equation}
In this section, we focused on constructing a suitable CVP instance,
\ie the lattice \(B\) and vector \(\vec{u}\).

\subsection{DSA Endgame: Lattice Attack} \label{sec:lattice_dsa}

In \autoref{sec:attack_dsa} we analyzed a timing vulnerability and then we collected
traces containing a variable amount of bits leaked during DSA signature
generation---we now show the lattice formalization enabling private key recovery.

\Paragraph{Lattice construction}%
Assume we have a sample of size \(N\) and elements of the form \((\omega_i,
r_i, s_i, h_i)\) where \(\omega_i\) is the elapsed time of signing a message
with hash \(h_i\) and signature \((r_i,s_i)\) satisfying \autoref{eq:rs-sign}.
Next, one proceeds by sorting the sample according to \(\omega_i\) and
retaining the \(f\ll N\) fastest ones, expected to correspond with
shorter-than-average nonces. After filtering, the next step is to
construct a suitable lattice that allows private key recovery.

Recall, in the filtered sample each nonce \(k_i\) \textit{should} have bit
length smaller than or equal to
\(m = \lg(q)\), and then \(k_i < q / 2^{\ell_i}\) for some positive integer \(\ell_i\).
Moreover, the inequality
$h_i / s_i + \alpha \cdot r_i / s_i \bmod q = k_i < q / 2^{\ell_i}$
is crucial because it ensures the existence of integers
\(\lambda_i\in\{-q \twodots q\}\) such that
\(\alpha \cdot (2W_i \cdot t_i) - (2W_i\cdot\hat{u}_i + q) - (2W_i\cdot\lambda_i\cdot q) \leq q\)
where \(t_i = r_i / s_i \bmod q\),
\(\hat{u}_i = -h_i / s_i \bmod q\),
and \(W_i = 2^{\ell_i}\).
To be more precise, with \(d \ll N\) a positive integer, the
dimensional-\((d+1)\) lattice
\begin{equation}
        B = 
        \begin{bmatrix}
                2W_1\cdot q       &       0               &       \cdots  &       \cdots          &       0       \\
                0               &       2W_2\cdot q       &       \ddots  &       \cdots          &       \vdots  \\
                \vdots          &       \ddots          &       \ddots  &       0               &       \vdots  \\
                0               &       \cdots          &       0       &       2W_d\cdot q       &       0       \\
                2W_1\cdot t_1     &       2W_2\cdot t_2     & \cdots        &       2W_d\cdot t_d     &       1
        \end{bmatrix}
	\label{eq:lattice:B}
\end{equation}
and the integer vectors \(\vec{u} = (2W_1\cdot\hat{u}_1 + q, \ldots, 2W_d\cdot\hat{u}_d + q, 0)\), 
\(\vec{z} = (\lambda_1, \ldots, \lambda_d, \alpha)\), and \(\vec{y} = (2W_1\cdot \nu_1, \ldots, 2W_d\cdot \nu_d, \alpha)\) 
satisfy \(\vec{z}B - \vec{u} = \vec{y}\) with
\(\nu_i\in \{-(q-1) / 2 \twodots (q-1) / 2 \}\)
the signed modular reduction of
\(\hat{u}_i + q / (2W_i) \bmod q\).

\Paragraph{Lattice parameters}%
Feasible private key recovery depends on lattice parameters \(N\), \(f\), and \(d\).
Recall, \(N\) determines the number of measured signatures, and each signature requires a nonce \(k_i\)
randomly and uniformly drawn from \(\{1 \twodots q-1\}\). Notice the expected number of nonces with \(\ell\)
MSBs clear is approximately \(N / 2^\ell\). Our goal is to determine the
expected number of MSBs clear for each nonce corresponding with the filtered sample size \(f\).

Denote \(\ell\) the number of clear MSBs for the dimensional-\((d+1)\) lattice
construction. Each column of the lattice has \(\ell\) correlated bits to the private key and, assuming the private
key has \(m\) bits, the expected number of independent bits in a sub-sample of size \(d\) is 
\(m \cdot (1 - (1 - \ell / m)^d)\)~\footnote{Each correlation can be viewed as an \(\ell\)-tuple
with entries determined by the bit positions of the private key that are correlated. Thus, the number of different positions
in a sample of \(d\) tuples with entries in \(\{0 \twodots m -1\}\) will determine the number of independent bits.}.
Then setting \(d = c\cdot m / \ell\), the probability that a random sub-sample of size \(d\) has \(m\) independent
correlated bits is
$1 - (1 - \ell / m)^d \approx 1 - e^{-c}$
and \(c = 1.25\) is enough to ensure a ``small'' lattice dimension and success probability of at most \(0.71\).
Hence we set \(c = 1.25\) in what follows.

\begin{figure}
\iftrue
\includegraphics[width=\linewidth]{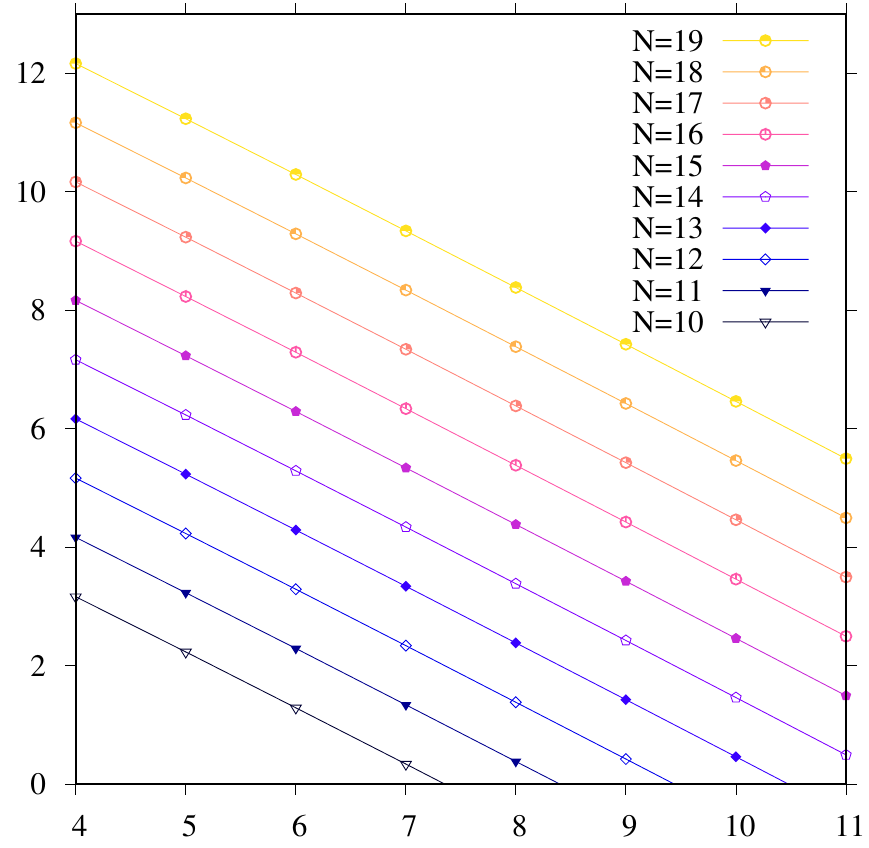}
\else
\Huge\bf PLACE\\HOLDER
\fi
\caption{The x-axis corresponds with the number \(\ell\) clear MSBs used, while
the y-axis determines the quantity \(\theta - \ell\)
where \(\theta = \lg(N) + \lg(\ell) - \lg(m) - \lg(1.25\cdot\delta)\)
is the expected number of clear MSBs in the filtered sample.
Sample sizes are log scale.}
\label{fig:ell}
\end{figure}

Increasing \(f\) decreases the expected number of clear MSBs, hence the best
configuration is to set \(f = \delta \cdot d \approx d\) with
\(\delta \in (1,2]\) and \(f = N / 2^\theta\) where
\(\theta = \lg(N) + \lg(\ell) - \lg(m) - \lg(1.25\cdot\delta)\).
Finally, the existence of noise in the measurements necessitates a certain
degree of freedom for the number of clear MSBs. 
Recall, \(\theta\) is the expected number of clear MSBs (for noise-free measurements)
and \(\ell\) is the fixed number of clear MSB's to be used, and therefore the quantity 
\(\theta - \ell\) gives an idea of how much large \(\ell\) can be for a fixed sample size \(N\). 
\autoref{fig:ell} illustrates
the degree of freedom assuming \(\ell \in \{4 \twodots 11\}\) and different
sample sizes.
\autoref{sec:lattice_work} shows experiment results for the
lattice applied to our collected samples during the remote timing attack scenario.

\subsection{ECDSA Endgame: Lattice Attack}\label{sec:lattice_wnaf}
This section describes two kinds of lattice constructions, depending on
the SCA nature.
\textit{Unsigned} applies to signals where point subtractions cannot be
distinguished from point additions during wNAF point multiplication (\cf \autoref{sec:attack_wnaf}).
\textit{Signed} is a stronger attack in the sense that point subtractions
\textit{can} be distinguished, yielding the sign of each wNAF representation
coefficient (\cf \autoref{sec:attack_sgx}). The lattice constructions utilizing said leakage are based on those
of \citet{DBLP:conf/ctrsa/PolSY15} and \citet{DBLP:conf/acsac/AllanBFPY16}, respectively,
summarized as follows.
With \wnaf window width
$w$, knowing the position of two consecutive non-zero coefficients \(\kappa_j\)
and \(\kappa_{j + \ell}\) leads to an equation with \(z\) $\alpha$-correlated
bits. Moreover, each column of the lattice \(B\) is determined by the pair
\((\kappa_j, \kappa_{j+\ell})\) along with public values \(h\) and \((r,s)\).

In addition, \autoref{eq:lattice:B} describes \(B\) but having
\(t = r \cdot s^{-1} \cdot 2^{m - j - \ell - 1}\bmod q\), \(W = 2^z\) where \(z = \ell - w\) and \(z = \ell - w + 1\) for
the unsigned and signed approaches, respectively.
In both approaches, \(\vec{u}\) has the same shape like in the timing attack but without the term \(q\), and either:
(i) \(\tilde{u} = 2^{m + w - \ell - 1} - (h\cdot s^{-1} + 2^{j + w} - 2^{j + \ell})\cdot 2^{m - j - \ell - 1} \bmod q\) for the unsigned case; or
(ii) \(\tilde{u} = (2b + 1)\cdot 2^{m + w - \ell - 2} - (h\cdot s^{-1} \cdot 2^{m - j - \ell - 1}) \bmod q\),
where \(b\) denotes the sign of the coefficient \(\kappa_{j + \ell}\) for the signed case.

\Paragraph{Remark}
The above equations are assuming each coefficient \(\kappa_j\) in the wNAF
representation of \(k\) satisfies \(-2^{w} < \kappa_j < 2^{w}\).
However, our case study uses a modified wNAF representation such that
\(-2^{w - 1} < \kappa_j < 2^{w - 1}\); that is, replacing \(w\)
with \(w - 1\).

\subsection{Lattices at Work} \label{sec:lattice_work}
To illustrate the practical implications of SVP instances used for private key
recovery, we implemented our lattice-based cryptanalysis in Python 3. After
constructing the dimensional-\((d+2)\) lattice \(\hat{B}\) from \autoref{eq:lattice:svp}, we proceed
by applying the method by \citet{DBLP:conf/eurocrypt/GamaNR10}. The main idea is
to reduce (using BKZ) another dimensional-\((d+2)\) lattice \(\tilde{B}\) which is computed by
shuffling the columns of \(\hat{B}\) and multiplying it by a unimodular matrix
with low density approximately equal to \((d + \sqrt{d})\). The goal is looking for a short lattice basis vector in the
reduced lattice of \(\tilde{B}\): if private key recovery is unsuccessful,
update \(\hat{B}\) as the reduced matrix of \(\tilde{B}\) and repeat the
procedure. Finally, our implementation solves the corresponding SVP instance
with help of the BKZ reduction included in
\code{fpylll}\footurl{https://github.com/fplll/fpylll}, a Python wrapper for
the \code{fplll} C++ library \cite{fplll}.

We focused our experiments on private key recovery for DSA and ECDSA procedures
with a 224-bit and \(384\)-bit \(q\), respectively. Each
experiment consisted of the following steps:
(i) select a random sample of size \(N\) from the SCA data;
(ii) construct a random dimensional-\((d + 2)\) lattice \(\hat{B}\);
(iii) look for a short lattice basis vector in \(\hat{B}\) allowing private key recovery;
goto (ii) if unsuccessful.

We labeled an experiment \textit{successful} when recovering the private key
according to the above steps and fixing the maximum lattice constructions to
correspond to roughly 4 h of wall clock single core CPU time.
The experiments with 224-bit DSA instances assume \(\ell = 4\) clear MSBs and use the suitable lattice dimension proposed in~\autoref{sec:lattice_dsa}.
On the another hand, experiments corresponding with 384-bit ECDSA have a different lattice nature not determined in terms of clear MSBs but by the distances between 
two non-zero wNAF coefficients. However, our lattice dimension choices are based on the analysis given in~\autoref{sec:lattice_dsa} and, because of the reduced sample size compared to
the DSA case, we increased the lattice dimension until reaching a (possible) high success probability.
The signed wNAF trace approach ensures one more
leakage bit and smaller sample sizes with a higher success probability than the (unsigned) wNAF trace approach. For both 224-bit DSA and 384-bit ECDSA, the small
density of about \((d + \sqrt{d})\) permits (at each step in the method by \citet{DBLP:conf/eurocrypt/GamaNR10}) the use of a ``random'' lattice with small difference
from the reduced one, and thus the BKZ reduction cost is minimized. In other words, we observe our choice \((d + \sqrt{d})\) allows a large number of different lattices in a fixed yet reasonable
amount of time, and therefore the success probability increase.

Additionally, the random samples used in 224-bit DSA instances correspond with the data obtained by the remote timing attacks presented in~\autoref{sec:attack_dsa}.
Conversely, the random samples used in the (unsigned) wNAF traces of 384-bit ECDSA instances are the ones obtained by the EM-based analysis presented in~\autoref{sec:attack_wnaf}.
As far as we known, this is first application of multi-digit lattice methods to EM-based signals.
And finally, the random samples corresponding to signed wNAF traces of 384-bit ECDSA instances
were obtained by the analysis from~\autoref{sec:attack_sgx}.

To have a better understanding how practice meets theory, we measured the elapsed
time and number of random lattices required in each successful experiment.
\autoref{tb:lattice} summarizes the average elapsed time (in minutes) and
the portion of successful experiments over all 1000 runs. Additionally, \autoref{tb:lattice}
also ilustrates the minimum, maximum, median, mean, and standard deviation of the number 
of lattice constructions performed. Moreover, from \autoref{tb:lattice} we can see that our improvements 
bring us a success probability of 0.38 (with sample size 1152) for 224-bit DSA, 0.14 and 0.83 (both with
sample size 30) for wNAF traces and wNAF signed traces of 384-bit ECDSA respectively.

\begin{table}[!bth]
        \resizebox{0.47\textwidth}{!}{%
        \begin{tabular}{|c|r|r||r|r|r|r|r||r|r|}
                \hline
                \textbf{Attack}  &
                \(N\)\phantom{Z}   &
                \(d+2\) &
                \textbf{Min}     &
                \textbf{Max}     &
                \textbf{Median}  &
                \textbf{Mean}    &
                \textbf{Stdev}   &
                \textbf{Time}    &
                \textbf{Ratio}   \\
                \hline \hline
                & 2048 & \multirow{4}{*}{78}
                & 1 & 1374 & 25 & 59.9 & 102.4
                & 0.3 & 0.99  \\
                Timing & 1536 &
                & 1 & 4826  & 6 & 48.9 & 255.7
                & 0.1 & 0.99 \\
                (\autoref{sec:attack_dsa}) & 1280 &
                & 1 & 10169 & 6 & 67.4 & 450.1
                & 0.1 & 0.74 \\
                & 1152 &
                & 1 & 10726 & 7 & 148.3 & 839.9
                & 0.1 & 0.38 \\
                \hline
                & 40 & 92
                & 1 & 2522 & 77 & 273.5 & 492.4
                & 2.1 & 0.11  \\
                & 40 & 132
                & 1 & 50 & 2 & 4.9 & 7.9
                & 0.9 & 0.09 \\
                wNAF & 40 & 172
                & 1 & 158 & 2 & 7.1 & 19.4
                & 1.3 & 0.08 \\
                (\autoref{sec:attack_wnaf}) & 30 &  92
                & 5 & 3673 & 522 & 781.4 & 872.7
                & 14.5 & 0.05 \\
                & 30 & 132
                & 1 & 1855 & 130 & 306.3 & 407.1
                & 9.8 & 0.14 \\
                & 30 & 172
                & 1 & 2008 & 173 & 322.6 & 425.0
                & 22.1 & 0.13 \\
                \hline
                & 40 & 92
                & 1 & 2725 & 35 & 197.4 & 425.4
                & 1.0 & 0.92  \\
                wNAF, & 30 &  92
                & 1 & 2814 & 357 & 641.9 & 711.6
                & 7.7 & 0.23 \\
                error-free & 30 & 132
                & 1  & 1969 & 178 & 370.5 & 442.8
                & 13.6 & 0.73 \\
                (\autoref{sec:attack_wnaf}) & 20 & 132
                & 47 & 1616 & 646 & 671.8 & 524.8
                & 47.5 & 0.02 \\
                & 20 & 172
                & 2  & 1617 & 771 & 792.6 & 517.1
                & 97.5 & 0.02 \\
                \hline
                & 40 & 92
                & 1 & 2817 & 3 & 80.2 & 277.4
                & 0.3 & 0.94  \\
                wNAF, & 30 &  92
                & 1 & 2854 & 101 & 430.5 & 672.6
                &  2.4 & 0.44 \\
                signed & 30 & 132
                & 1 & 840  & 13  & 76.2  & 145.1
                &  1.9 & 0.83 \\
                (\autoref{sec:attack_sgx}) & 20 & 132
                & 1 & 1893 & 363 & 569.3 & 622.3
                & 18.5 & 0.03 \\
                & 20 & 172
                & 4 & 2127 & 663 & 782.7 & 643.7
                & 64.2 & 0.04 \\
                \hline
        \end{tabular}
        }
        \caption{Minimum, maximum, median, mean, and standard deviation of the number of lattice constructions performed. Timings are in minutes, and correspond with the median of the successful experiments.}
        \label{tb:lattice}
\end{table}
\section{Conclusion} \label{sec:conclusion}

In this work, we presented an extensive SCA security evaluation of
Mozilla's NSS---from discovering vulnerabilities to performing key recovery attacks.
To identify potential SCA leaks, we combined an automated leakage assessment framework DATA
and the Triggerflow tool to identify the resulting leaks using during NSS invocations of DSA,
ECDSA and RSA using DATA and track the code paths that lead to them using Triggerflow.
The results led to the discovery of some serious SCA security flaws in
DSA and ECDSA nonce padding, ECDSA point multiplication and scalar encoding as well as RSA
key generation. To demonstrate real-world security impact, we also performed
several end-to-end attacks at the application level---remote timing attack on
DSA (research data released \cite{zenodo:2020:nss} in support of Open Science),
microarchitecture attack on ECDSA nonce encoding,
EM attack on ECDSA point multiplication and EM attack on RSA key generation.
Finally, we summarized the results of different lattice formulations used during key
recovery phase of the attacks.
Interestingly, the discovered vulnerabilities are known to the
research community and previously reported across multiple vendors (\eg OpenSSL),
which highlights a gap in the practice of CVE coordination among peer vendors.

\Paragraph{Cui bono?}
\NSS is certainly neither the first nor last security library to fall prey to
SCA and failure to use constant-time implementations. Why is this a recurring
event? Who should be held culpable? We note the break, fix, break cycle benefits
several stakeholders due to perverse incentives---to mention a few:
(i) it keeps software engineers in demand since these libraries are not ``deploy and forget'';
(ii) it keeps security engineers in demand since there is a steady stream of security issues to assess and address;
(iii) it keeps security researchers busy with a perpetual flow of research topics to write papers about---including us.
During judgment, the ancient Romans inquired \textit{Cui bono?} or ``Who benefits?'' to
identify suspects. Perhaps the incomplete list of key players above in this
self-perpetuating meta-system is a good start.

\Paragraph{Mitigations}
During responsible disclosure to Mozilla, we made several
FOSS contributions to assist in mitigating these issues and testing the
fixes---all of which are now merged.
(i) To solve the vulnerability in \autoref{sec:attack_dsa} (\CVE{2020-12399}), we proposed a
patch\footurl{https://hg.mozilla.org/projects/nss/rev/daa823a4a29bcef0fec33a379ec83857429aea2e}
to \NSS that randomizes the nonce by $\hat{k} = k + b \cdot q$ where $b$ is a
random word with a fixed bit length, \ie top bit set. Our empirical evaluation
of the patch indicates this aligns the curves in \autoref{fig:dsa_remote},
mitigating the issue.
(ii) For the vulnerability in \autoref{sec:attack_rsa} (\CVE{2020-12402}), we
implemented\footurl{https://hg.mozilla.org/projects/nss/rev/699541a7793bbe9b20f1d73dc49e25c6054aa4c1}
the constant-time GCD and modular inversion by \citet{DBLP:journals/tches/BernsteinY19}.
(iii) For the \autoref{sec:attack_wnaf} and \autoref{sec:attack_sgx}
vulnerabilities (\CVE{2020-6829}), we decided against patching \wnaf and,
similar to the \ecc{secp256r1} code in \NSS, proposed two custom \code{ECGroupStr} for
\ecc{secp384r1}\footurl{https://hg.mozilla.org/projects/nss/rev/d19a3cd451bbf9602672fdbba8d6a817a55bfc69} and
\ecc{secp521r1}\footurl{https://hg.mozilla.org/projects/nss/rev/ca068f5b5c176c503ddce969e78dd326cc5fd29a}.
We leveraged ECCKiila\footurl{https://gitlab.com/nisec/ecckiila} for this task \cite{temp:ecckiila},
built on top of \code{fiat-crypto}\footurl{https://github.com/mit-plv/fiat-crypto}
to take advantage of its formally verified and constant-time GF layer \cite{DBLP:conf/sp/ErbsenPGSC19}.
(iv) With these constant-time versions in place and \NSS not featuring any
other vulnerable curves, the broken fix in \autoref{sec:attack_padding} (\CVE{2020-12401}) is no
longer needed---hence we submitted a patch to remove the
padding\footurl{https://hg.mozilla.org/projects/nss/rev/aeb2e583ee957a699d949009c7ba37af76515c20}.

\Paragraph{Acknowledgments}
We thank Tampere Center for Scientific Computing (TCSC) for generously granting
us access to computing cluster resources.

The first author was supported in part by the Tuula and Yrj\"o Neuvo Fund
through the Industrial Research Fund at Tampere University of Technology.

This project has received funding from the European Research Council (ERC) under
the European Union's Horizon 2020 research and innovation programme (grant
agreement No 804476).
 
\bibliographystyle{ACM-Reference-Format}

\ifAPPENDIX
\clearpage
\appendix
\section{Reference Algorithms}
\begin{algorithm}[h]
	\caption{Compute \wnaf representation of $k$}\label{alg:compute_wanf}
	\DontPrintSemicolon
	\KwIn{Integer $k$ and width $w$}
	\KwOut{$wNAF(k,w)$}
	\SetKw{KwDownTo}{downto}
	\SetKwFunction{odd}{odd}
	$i=0$\\
	\While{$k \ne 0$}
	{
		\If{$\odd(k)$}{
			$d = k \bmod 2^w$\\
			\lIf{$d > 2^{w-1}$}
			{
				$d = d - 2^w$
			}
			$k = k - d$
		}
		\lElse
		{
			$d = 0$
		}
		$wNAF[i] = d$, $k = k / 2$, $i = i + 1$
	}
	\Return{$wNAF$}
\end{algorithm}
\begin{algorithm}[h]
	\caption{\wnaf-based scalar multiplication}\label{alg:wnaf_scalar_mult}
	\DontPrintSemicolon
	\KwIn{Integer $k$, width $w$ and elliptic curve point $G$}
	\KwOut{$R=kG$}
	\SetKw{KwDownTo}{downto}
	Compute $wNAF(k,w)$ using \autoref{alg:compute_wanf}\\
	$P[i] = iG$ ; $i \in \{-2^{w-1} + 1, \twodots, -3, -1, 1,  \twodots, 2^{w-1} - 1\}$\\
	$R = \Oinf$\\
	\For{$i = \lfloor \lg(k) \rfloor + 1$ \KwDownTo $0$}{
		$R = 2R$ \\
		\lIf{$wNAF[i] \ne 0$}{%
			$R = R + P[wNAF[i]]$}%
	}
	\Return{$R$}
\end{algorithm}
\begin{algorithm}[h]
\caption{Binary extended Euclidean algorithm (BEEA)}
\label{alg:BEEA_alg}
\DontPrintSemicolon
\KwIn{Integers $a$ and $b$ such that $0<a<b$}
\KwOut{Greatest common divisor of $a$ and $b$}
\SetKwFunction{even}{even}
\Begin{
    $u \gets a$, $v \gets b$, $i \gets 0$\;
    \While{\even{$u$} {\bf and} \even{$v$}}{
        $u \gets u / 2$,
        $v \gets v / 2$,
        $i \gets i + 1$
    }
    \While{$u \ne 0$}{
        \lWhile{\even{$u$}}{
            $u \gets u / 2$
        }
        \lWhile{\even{$v$}}{
            $v \gets v / 2$\
        }
        \lIf { $u \ge v$}{
            $u \gets u - v$
        }\lElse{
            $v \gets v - u$
        }
    }
    \Return{$v \cdot 2^i$} \label{algstep:gcd_return}
}
\end{algorithm}
 \fi

\end{document}